\documentclass[11pt,letterpaper]{article}

\usepackage{amsmath}
\usepackage{amssymb}
\usepackage{graphicx} 
\usepackage{mathabx} 
\usepackage{natbib}
\usepackage{enumitem}
\usepackage[]{subcaption}  
\usepackage[caption=false]{subfig}  
\usepackage{authblk} 
\usepackage{deluxetable}
\usepackage{setspace}  

\setlist{nolistsep} 

\ifx\UseOption\undefined
\def\UseOption{optb}
\fi
\usepackage{optional}
\providecommand{\e}[1]{\ensuremath{\times 10^{#1}}}

\begin{document}
\singlespacing 

\setcounter{page}{1}
\setcounter{section}{0}




\title{Iron Abundance Correlations and an Occurrence Distribution Discrepancy from Ongoing Planet Migration} 
%
\author{Stuart F. Taylor}
\affil{Participation Worldscope \\ Sedona, Arizona, USA \\ astrostuart@gmail.com}
\date{\vspace{-5ex}} 
\maketitle


\begin{abstract}
Whether the higher occurrence of giant planets being hosted by 
metal-rich versus metal-poor stars results from 
formation or from ``pollution'' has been a question of intense debate.
We present new patterns that emerge when   planet/star systems are separated by 
stellar [Fe/H], and when systems with stellar companions are separated out. 
These differences can best
be explained if the onset of high eccentricity planet migration is 
also a time when planet are sent into merge with the star. 
Planet migration into the star is likely a 
complementary explanation to the view that systems with higher initial iron abundance
form more planets, and that more crowded planets are more likely to scatter into eccentric orbits.
Planets of iron-rich single stars have eccentricity distributions that are higher 
than planets of iron-poor single stars 
(where ``rich'' and ``poor'' are stars whose [Fe/H] is above and below solar, respectively). 
Stars with planets that have a stellar companion  comprise a third population of 
systems in which the stars are preferentially iron-rich. 
We describe new patterns that 
are best explained by eccentric planet 
migration being associated with other planets migrating into the star.
Though medium planets are more numerous than giant planets at periods greater than
three days, giant planets are more numerous than medium planets at the shortest periods.
Since giant planets migrate into the star faster, we show this as evidence of giant planet migration.
Planet migration into the star is certain to be an important part of 
planetary system evolution.
\end{abstract}


\section{Introduction}
\label{sec:intro}                
\subsection{Formation versus accretion}
\label{ssec:debateFvsA}                

Ever since it was found that stars with higher iron abundance are more likely to host planets, 
there has been a debate over whether this was due to more planets being formed around stars 
with more iron, or whether the stars have been polluted by iron rich material accreted during or 
after the planet forming process. The abundance patterns of elements other than iron have been 
used to argue that accretion from the planet forming disk would produce different patterns of
refractory versus volatile elements than those observed, leading authors to conclude that 
pollution was not a viable explanation. This, however, ignored that pollution may be due 
to whole planet infall, and that whole planet infall may continue to occur long after planet 
formation. Pollution has been assumed to remain in the surface convection zone for the life 
of the star, but it is not known if pollution may be slowly convected below, such that 
more recent pollution may leave higher iron abundances than primordial pollution.

Whether the correlation of 
higher planet occurrence with higher metallicity is
due to formation or pollution has been a long-running debate.
The hypothesis that higher metallicity is due to pollution has commonly been 
that the pollution is due to accretion of material from the disk, while
pollution due to whole-planet migration is less often considered.
Indeed, the pollution scenario has been dismissed based 
in part on the lack of a correlation with eccentricity and period
as predicted by the accretion of material  \citep[hereafter FV05]{fis05}.  

The presence of more 
shortest period giant planets than would be expected from tidal migration theory 
given the tidal migration strength calculated from binary star studies 
has also led to a debate \citep{ham09,hel09} 
over whether weak tidal dissipation in the star results in slow enough migration
for us to still find a few planets at such close distances \citep{pen12}, 
or if there is a resupply of these planets which we really are finding during the
time before they infall into the star \citep{tay12a,tay12b}.
 
The seminal discovery of Peg 51b in 1995 showed that planets
move from where they form more than anticipated.  The amount of 
planet migration is (arguably) the best studied surprise from the discovery of sufficient planets 
to do statistical studies of planet systems.  
Planet migration inevitably sends planets into the star \citep{mat13}.




\subsection{Metallicity correlation search history} 
\label{subsec:MetalHist} 


That higher metallicity stars have a higher occurrence rate has been known
since reported by \citet[hereafter G97]{gon97}, with
frequently cited confirmations including 
\citet[hereafter S03]{sant03} and \citep[hereafter FV05]{fis05}.
There were not enough of the highest eccentricity planets, and the presence
in the low iron abundance systems of a narrow period spike
of high eccentricity planets masked the appearance of the
[Fe/H]$_{\ast}$-eccentricity correlation at these earlier times, even if it was beginning to appear.
%

G97 pointed out that if the giant planet metallilcity correlation 
was due to ``pollution'' from accretion from the disk 
then the measured level of metallicity should increase with increasing mass of the star,
because more massive stars
decrease
in the depth of the convective zone (CZ)  mass.
G97 showed that pollution from grains would have an elemental 
dependence on condensation temperature, 
but the elemental concentration does not vary with condensation temperature, which was 
interpreted as evidence against pollution from the primordial disk. However,  
pollution from scattered planets would not have this variation with condensation temperature.
S03 looked for correlations of metalliticy with eccentricity, but  
at that time there were too few planet orbits with an eccentricity above 0.35
for this correlation to have been detected.
In retrospect, the plot in S03 does have the appearance that this correlation was beginning to appear for higher
eccentricities, but they did not yet have the statistics to confirm the trend. 
S03 report a possible paucity of metal-poor planets at the lowest eccentricities, 
but we find that subsequent data has found more planets in this region.

FV05 performed a thorough study of planet occurrence looking for many correlations.  
They report not seeing the trend in concentration of refractory elements 
expected if disk pollution contributed to metallicity.
A plot in FV05  vaguely shows 
that higher eccentricity systems were more likely among low [Fe/H]$_{\ast}$
systems, but at that time there were too few statistics to confirm this trend.
\citet{hal05} concluded with 79\% significance that metallicity is not correlated with eccentricity
based on an analysis of 72 SWPs. 

\citet[hereafter DM13]{daw13} found the correlation with eccentricity, and also found that 
in the Kepler data the three day pileup of giant planets is a feature of high [Fe/H]$_{\ast}$ systems.
Their explanation is that higher eccentricity results from higher rates of scattering of
giant planets, which scattering is a result of more crowded giant planet formation from
molecular clouds of higher metallicity. Having a higher rate of scattering would certainly
result in higher rates of planet infall. We find that the new correlations of [Fe/H]$_{\ast}$ are best
explained by a combination of higher metallicity leading to more crowded giant planet formation
followed by the [Fe/H]$_{\ast}$ being increased still further by planet infall.

We present correlations of [Fe/H]$_{\ast}$ in stars with giant planets (SWGPs) with the orbital 
eccentricity and the presence of a binary stellar companion, which we
found while searching for evidence of planets infalling and merging with stars.
We were motivated by evidence in the planet occurrence distributions 
that more giant planets than smaller planets
are migrating towards the star.
In Section~\ref{sec:signatures} we will show that 
this is better explained by the ``flow'' of planets suggested
by \citet{soc12} than by invoking weak enough tidal dissipation in the star
for the planets to remain suspended in such short-period orbits for times on the
order of the age of the system. It does not require unreasonable
numbers of giant planets migrating through a ``Socrates flow'' to say that
the tidal quality factor ``$Q^{\prime}_{\ast}$'' is not too different from the
$Q^{\prime}_{\ast}$ for stellar companions. Though there has been good
work proposing mechanisms for such a large dependence of  $Q^{\prime}_{\ast}$ on
mass, Section~\ref{sec:signatures} uses our and others' direct evidence of giant planet migration 
to show that the better explanation for the presence of these short-period planets
is planets migrating inward.

We describe newly found patterns that  distinguish iron-rich from iron-poor systems.
(1.) The eccentricity distribution of iron-poor systems has a 
maximum that rises with the period, up to a ``spike'' of high-eccentricity 
iron-poor systems that are mostly confined to a narrow period range.
Iron-rich systems have eccentricities distributed higher than iron-poor systems
at periods outside of this ``spike''.
(2.) The occurrence distribution of iron-rich single-star systems contains 
a gap of planets that starts at the same period range as the spike 
of high-eccentricity iron-poor systems. 
 
A significant part of the stellar [Fe/H]$_{\ast}$ 
 correlation with planet orbit eccentricity is best explained
 by saying that planets scattered by the eccentric planet migration inward ``pollute'' the star.
 If this pollution is only temporarily observed due to being mixed in the star, then
 this would explain why planets with higher eccentricities have higher [Fe/H]$_{\ast}$:
 stars with 
 the shortest period planets that have already been circularized, including
 those with spin orbit misalignment, may have already had more time for the
 star to have mixed the pollution away from the star. This would also explain
 why we find the correlation of stellar [Fe/H]$_{\ast}$ with eccentricity to be
 stronger for SWGPs of periods under 500 days than for SWGPs of 
 periods more than 600 days. The longer migration times of the longer period
 planets may have allowed more time for the pollution to mix away from the
 surface of the star. 
We cite the paucity of 2nd planets in systems with giant planets in orbits shorter
than three days to suggest that  other planets can be expected to have been
scattered into the star. We disagree that the high [Fe/H]$_{\ast}$ correlated with the presence
of giant planets must be due to pollution or scattering alone, rather we
find the hypothesis of \citet{daw13} that higher metalllicity molecular
clouds form more crowded giant planet systems that leads to the scattering
that forms the three-day pileup complementary to our hypothesis:
such a scenario of increased scattering could raise the [Fe/H]$_{\ast}$ still further by
pollution.

We also present a newly found correlation of the [Fe/H]$_{\ast}$ of SWPs with the presence of
a stellar companion in addition to the planet. This correlation challenges both
the formation and pollution explanation of higher [Fe/H]$_{\ast}$. Because stellar companions
are expected to be formed by gravitational instability, binary SWPs (BSWPs)
would not be expected to have higher [Fe/H]$_{\ast}$, so unless some stars are
seeded by core accretion, the primordial explanation of high [Fe/H]$_{\ast}$ in SWGPs 
fails to explain how a binary companion could be correlated with high [Fe/H]$_{\ast}$.
Unless a binary companion also corrals planets or material into the star,
the pollution scenario is also challenged.

\subsection{``Planet loss": Migration into and merger with the star}  
\label{ssec:planetloss}                
Planets not only migrate far more than expected, planets also 
continue to migrate after the standard migration that occurs due to interaction with
the planet-forming disk. The slope of the occurrence function of the shortest period in giant planets
shown in Section~\ref{sec:signatures} 
indicates that migration continues throughout the main sequence lifetime of the star.

Though it has long been known that [Fe/H]$_{\ast}$ in a star is correlated with hosting giant planets \citep{gon97}, 
Section~\ref{ssec:bswp} shows that the [Fe/H]$_{\ast}$ 
of a star is also correlated with the presence of a binary companion,
in addition to the recent finding that [Fe/H]$_{\ast}$ is correlated 
with high planet eccentricity \citep{tay12b,tay13a,daw13}.
The newly found correlation with a star being in a binary system,
as well as the correlation with orbital eccentricity, was found motivated by
how a companion could be associated with planet scattering  that could increase pollution of the star.

``Hot Jupiters'' were an unexpected surprise when the first such short period planet was found in 1995, 
but now the surprise is that there appear to be too many of these shortest period planets.
Most extremely short period planets are doomed to tidally migrate into their stars \citep{lev09,jac09,ham09}. 
\citet{ham09} points out how it is unlikely that we are seeing more than 
two planets in the last 1/1000 of their existence, as would be required 
if stars had the expected rate of tidal friction. 
\citet{pen12}, \citet{ham09}  and others  
have proposed that planets take a long time to migrate into the star due to
weak tidal friction raised by the planet on the star, 
saying that otherwise some of these planets would have to be 
perturbed from further away by ``unlikely'' events. However, \citep{tay12b,tay13a} 
show that the rate of planets that would need to flow in is reasonable, especially given recent results 
of a ``flow'' of high eccentricity giant planets migrating inwards
\citep{soc12}. 

A resupply of the shortest period 
giant planets could be reasonably accomplished with only 
on the order of  
$10^{-12}$ giant planets per star per year \citep{tay13a}.
These planets will migrate into the star at rates that the resulting transient events 
should be found by current or upcoming transient surveys, 
but are still low enough rates to explain why they have not regularly been seen by past surveys. 
\citet{tay12b,tay13a} 
presents that ``hot Jupiters'' may be the result of giant planets migrating inwards, 
scattering away other planets in the process, and leaving giant planets with higher eccentricities 
in systems with few additional planets.

The recent discovery of the correlation between increasing [Fe/H]$_{\ast}$ in the star 
and increasing eccentricity among planets \citep{tay12b,tay13a} was prompted by the 
hypothesis that the occurrence distribution evidences a flow of giant planets into the star. 
The search was also motivated by the observation that few hot Jupiter systems have 
additional planets beyond the hot Jupiter, so we were seeking to find evidence 
if the missing planets may have gone into the star.  The correlation was found when the search was restricted 
to planets with periods less than 500 days, motivated by 
the hypothesis that scattering from further out may not only have been less 
likely to scatter other planets into the star, and also that at longer periods 
high eccentricity migration would be much slower, but that perhaps
the pollution would be mixing away from the surface of the star during the time that
the planets migrated from high eccentricity orbits to short-period low eccentricity orbits.  
We found that for planets with eccentricities greater than 0.35 in this period range, [Fe/H]$_{\ast}$ clearly increases with eccentricity. We find a weaker correlation for planets with longer periods. This increase led us to consider that the pollution signal may get washed out of the surface layer of the star during the time it takes for migration to less eccentric orbits.




\subsection{Organization}
\label{subsec:organization}                
Planet migration towards the star, leading to planet star mergers, are discussed in
Section~\ref{sec:migmerge}
The correlations of iron in the star with the eccentricity of the planet orbit and
the presence of a binary companion are presented	 in Section~\ref{sec:correlation}.
We present the difference in the flow rates of giant and ``medium" 
(Neptune-size) planets in Section~\ref{sec:signatures}, 
where we describe how a flow of giant planets can reconcile this discrepancy
and at the same time remove having to say that planets induce
weaker tidal dissipation in stars than do stellar companions.
We give some motivation for this work and suggest what kind of
observations could check these results in Section~\ref{sec:discussion}, 
discuss what further work we will do in Section~\ref{sec:FurtherWork}, 
followed by the conclusions (Section~\ref{sec:conclusions}).

\begin{figure}[htb]  
\centering
  \subcaptionbox{All systems.\label{fig:EccVsPerSzFe-All}}
  {\includegraphics[width=0.48\textwidth]{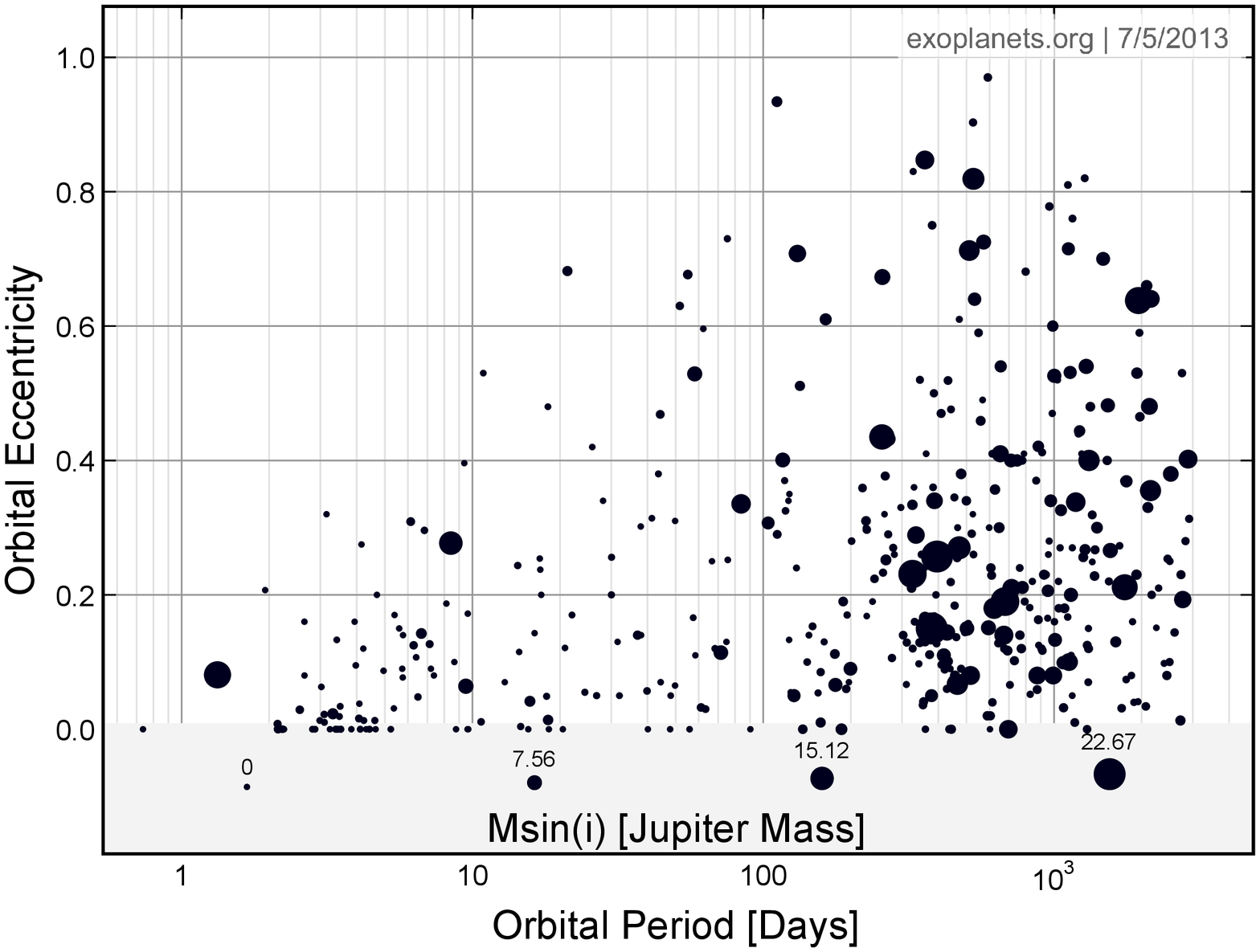}} 
  \subcaptionbox{Systems with [Fe/H]$_{\ast}$ $\le$ solar. \label{fig:EccVsPerSzFe-ltfe0}}
  {\includegraphics[width=0.48\textwidth]{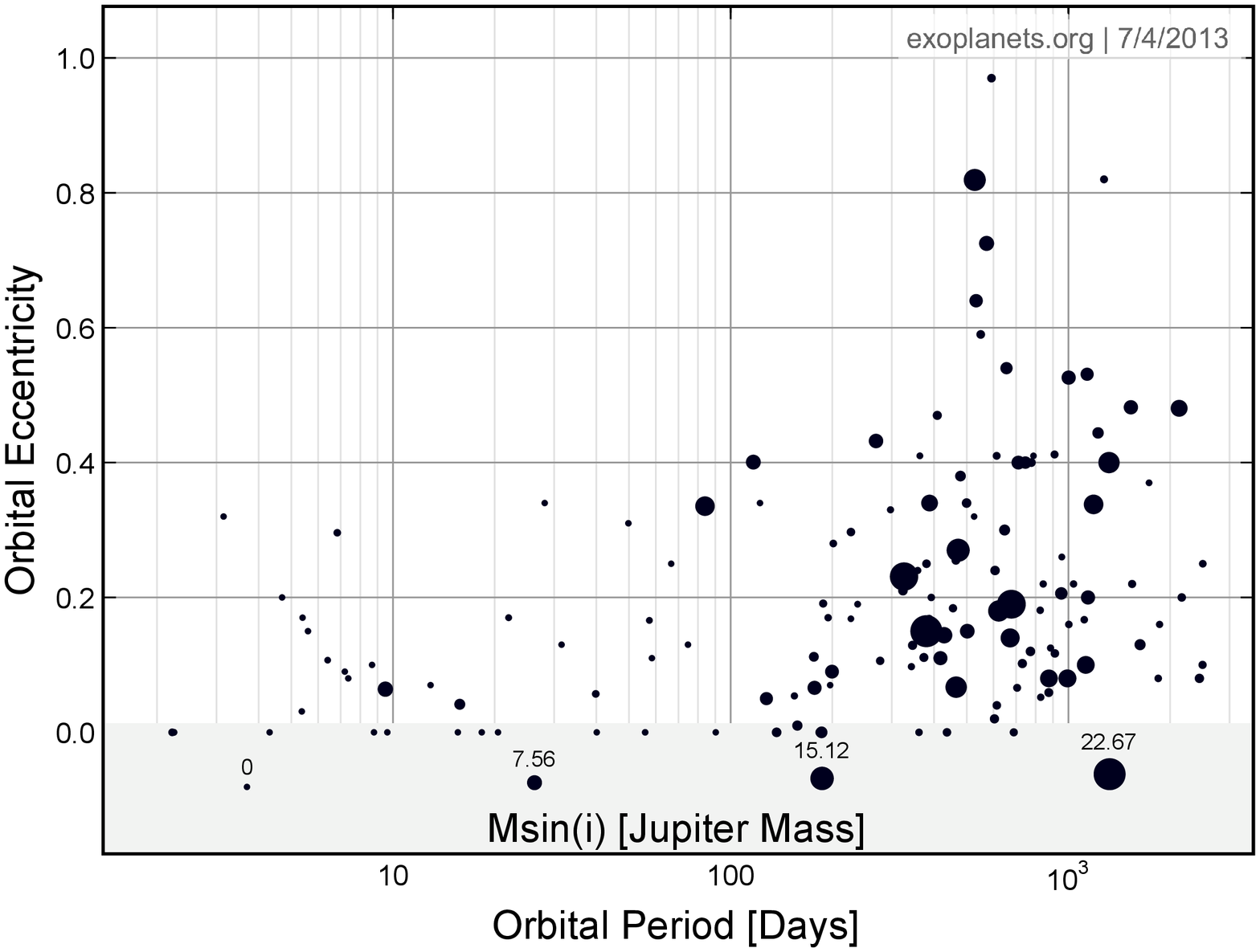}}
\caption{Eccentricity as function of period of planets found by radial velocity. 
  The symbol size indicates the radial component of the planet mass,  
  ($m_\mathrm{planet} \sin i$), in Jupiter masses ($M_{\rm J}$).
  While the entire sample (a) shows a distribution up to 
  high eccentricities, the iron poor systems (b) tend to have more circular orbits (e \textless 0.6) 
  except for a spike of a few systems with periods between 520 and 600 days. 
  The only iron-poor, high-eccentricity system outside this period range, 
  HD 7449b, may have its eccentricity excited by a distant companion of uncertain mass 
  \citep{dum11}.  
  }
  \label{fig:EccVsPerSzFe}
\end{figure}

\begin{figure}[htb]  
\centering
  \subcaptionbox{All systems discoverd by RV.\label{fig:PerHistFeHiVsLo-a-aswp}}
  {\includegraphics[width=0.48\textwidth]{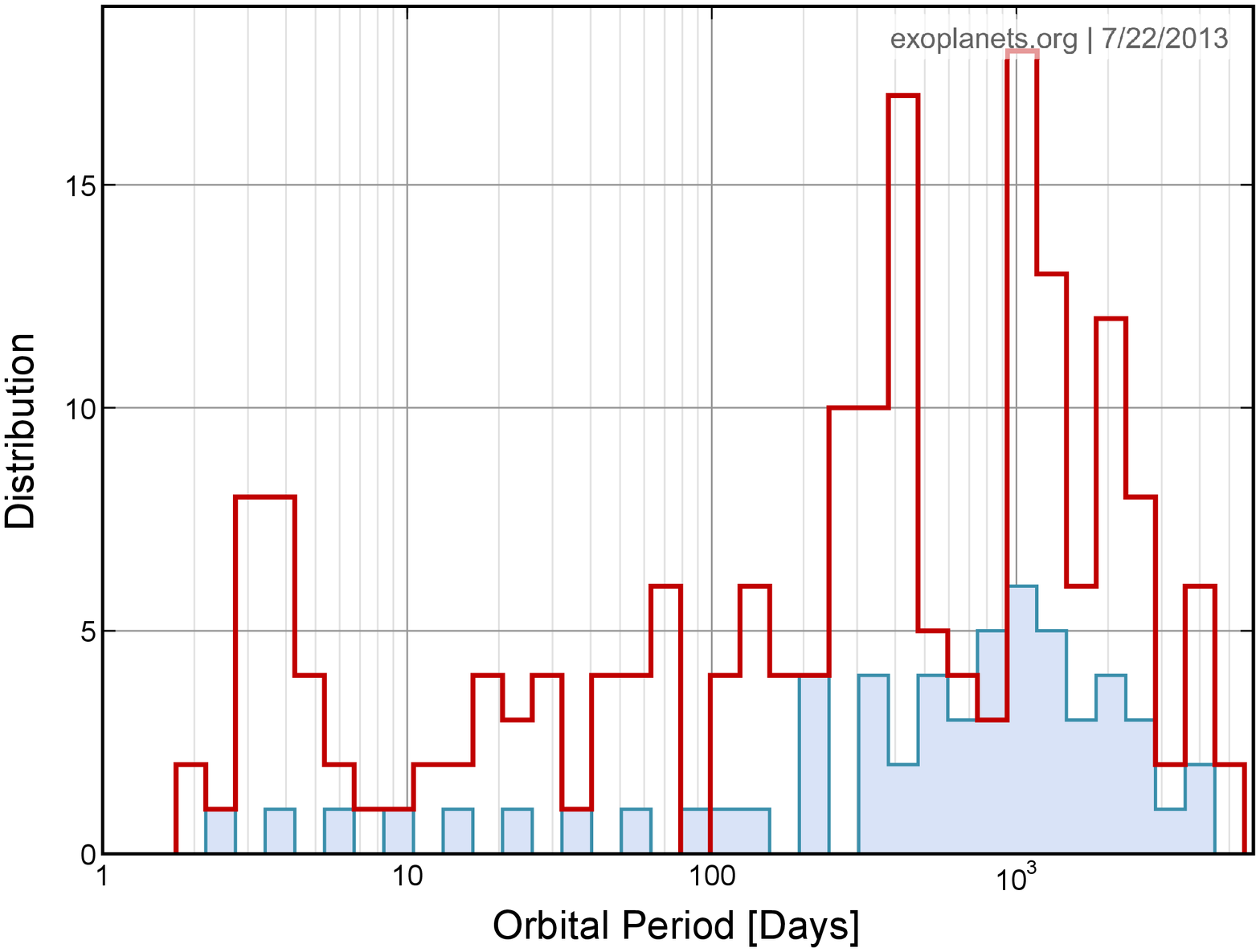}} 
  \subcaptionbox{Only single-star systems \label{fig:PerHistFeHiVsLo-b-sswp}}
  {\includegraphics[width=0.48\textwidth]{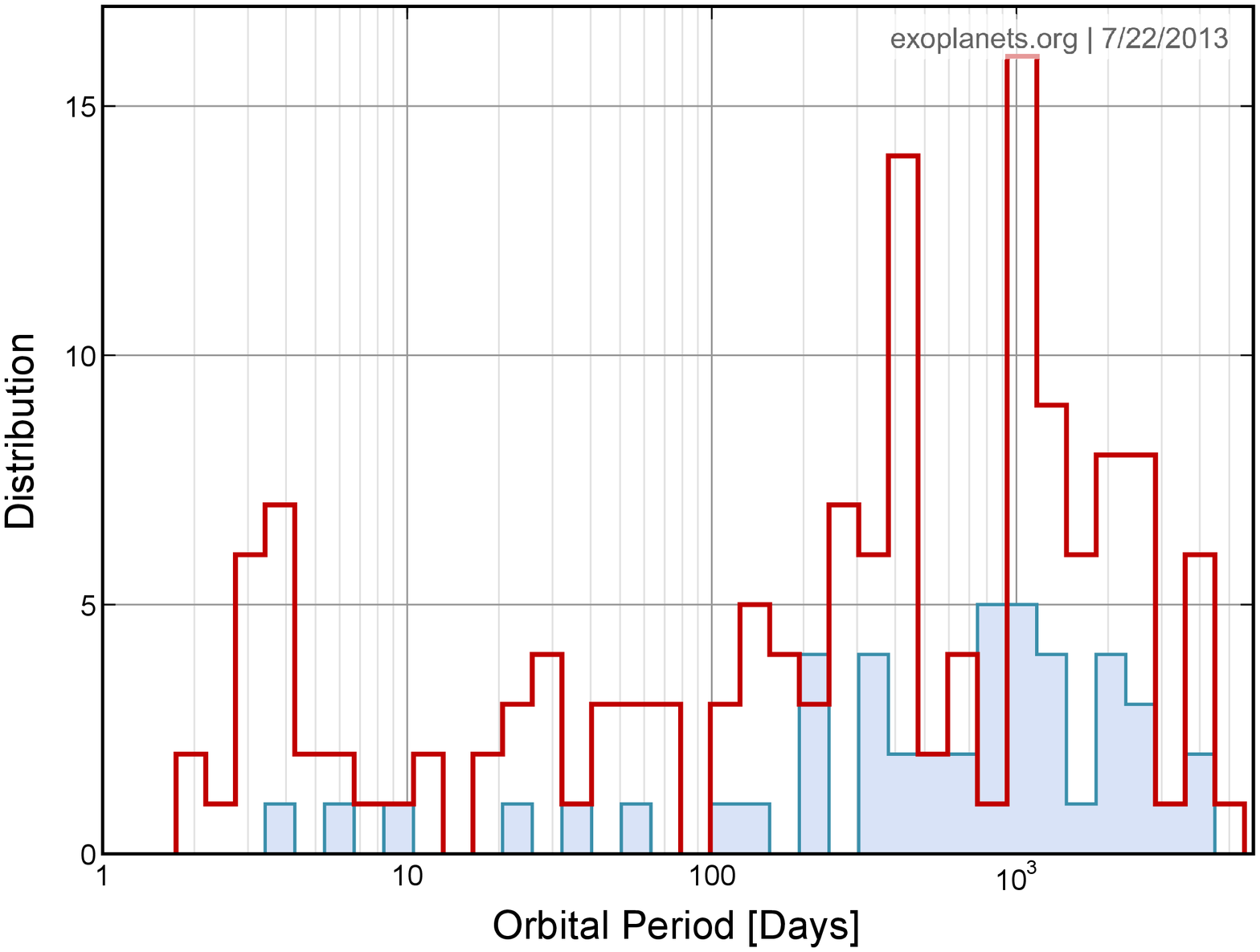}}
\caption{ The distribution by period of systems found by radial velocity that have 
$m_\mathrm{planet} \sin i \geq 0.1 M_\mathrm{Jupiter}$, 
separately counting planets hosted by iron-rich stars ([Fe/H]$_{\ast}$ \textgreater 0, red heavy line, unfilled),
and iron-poor stars  ([Fe/H]$_{\ast}$ $\leq 0$, blue filled bins).
These histograms show that while iron-poor systems have a single broad peak 
centered at approximately 1000 days, that iron rich systems have a two-peak
distribution separated by a gap 
going from under 500 days to over 900 days. 
We show this with (\protect\subref{fig:PerHistFeHiVsLo-a-aswp})
and without (\protect\subref{fig:PerHistFeHiVsLo-b-sswp})
including systems that have a stellar companion, 
because since binary star systems are being showne here to have higher [Fe/H]$_{\ast}$, their distribution may be different.
The gap may indeed be deeper when only single star systems are considered, 
(\protect\subref{fig:PerHistFeHiVsLo-b-sswp}).
  }
  \label{fig:PerHistFeHiVsLo}
\end{figure}

\begin{figure}[h] 
        \begin{subfigure}[ht]{0.48\textwidth}
                \centering
                \includegraphics[width=\textwidth]{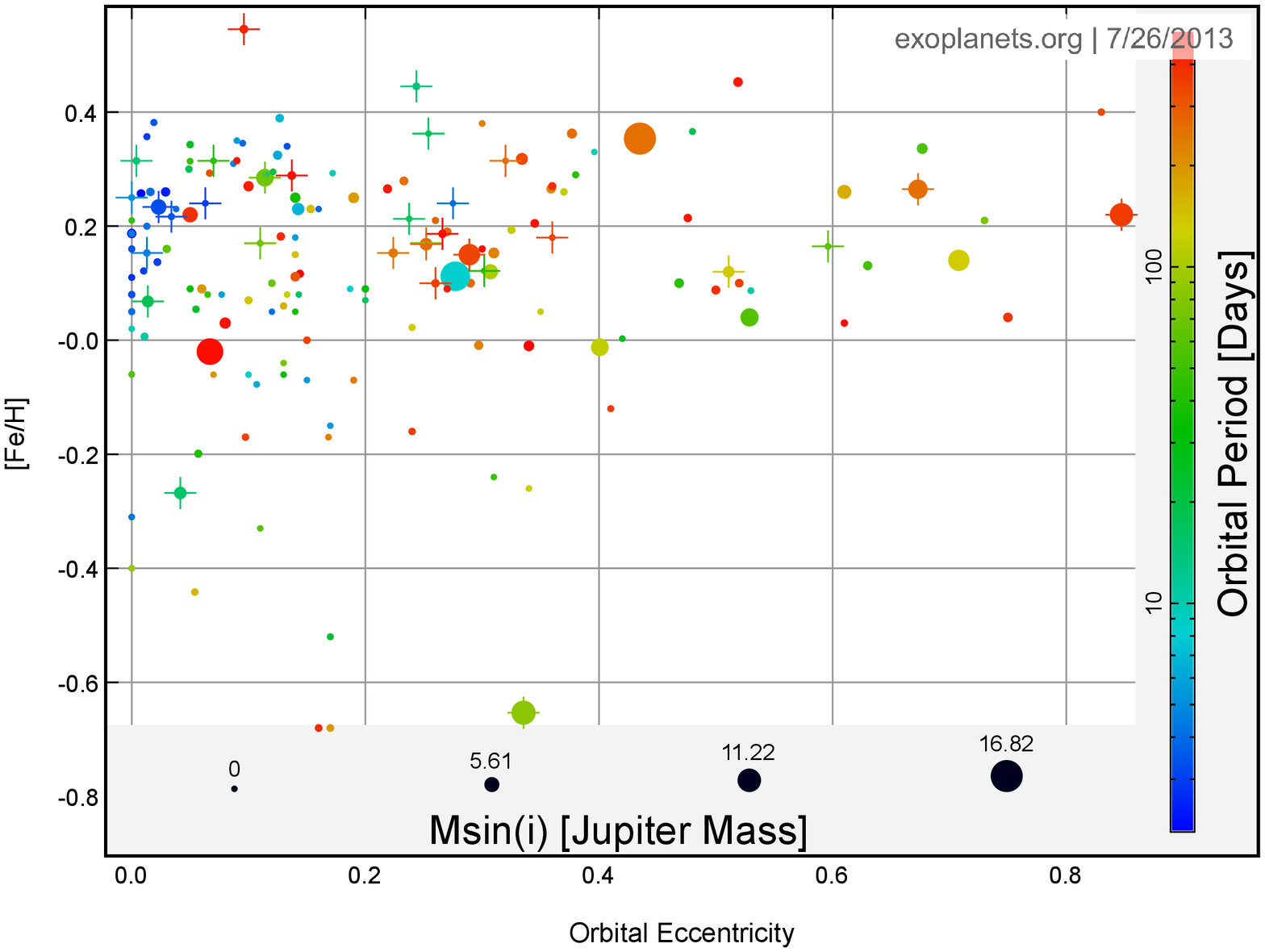}
                \caption{Periods less than 500 days, }
                \label{fig:fehplt500}
        \end{subfigure}%
        \begin{subfigure}[ht]{0.48\textwidth}
                \centering
                \includegraphics[width=\textwidth]{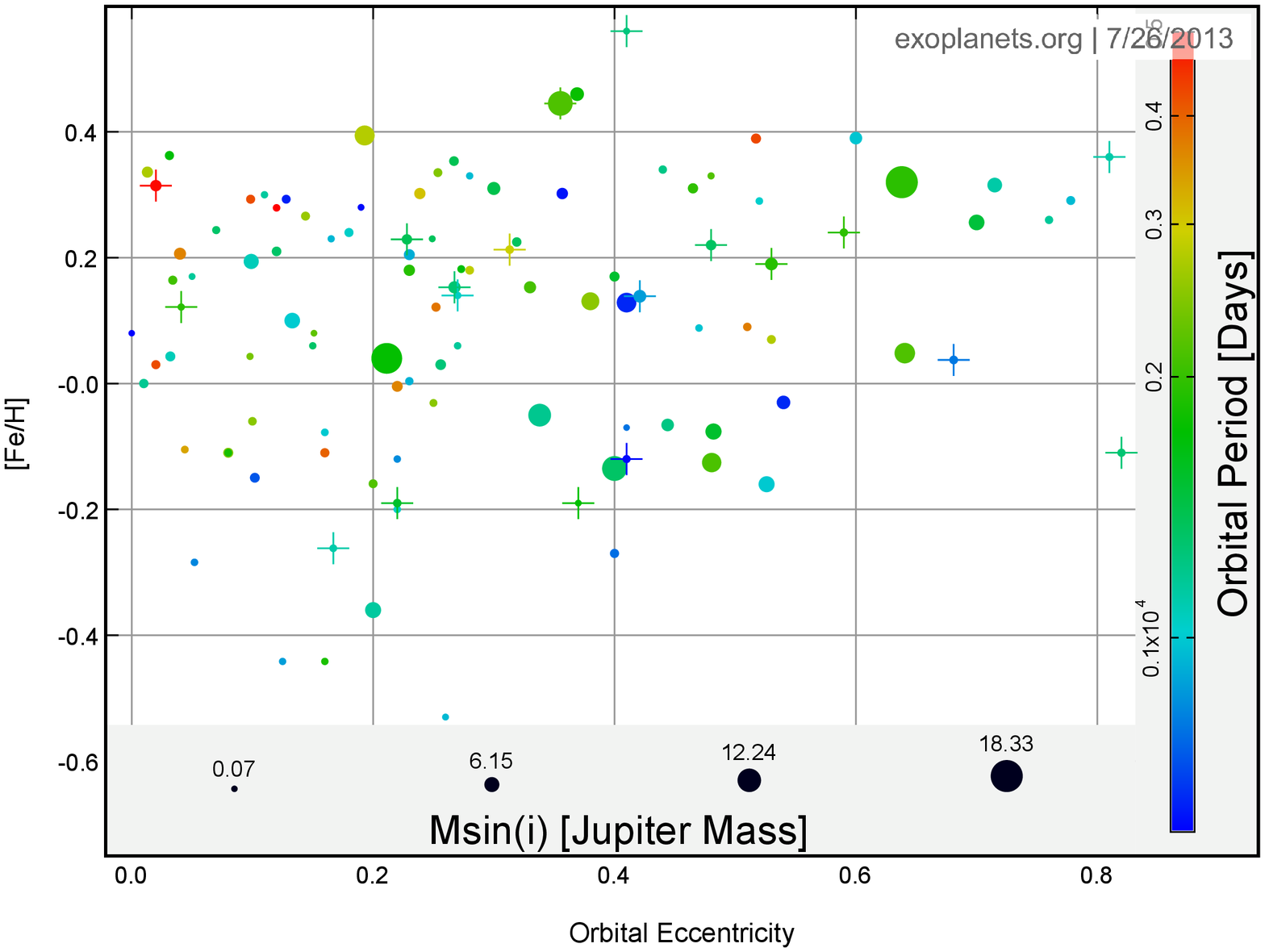}
                \caption{and greater than 600 days.}
                \label{fig:fehpgt600}
        \end{subfigure}
\caption{
  [Fe/H]$_{\ast}$ versus orbital eccentricity for (\protect\subref{fig:fehplt500}) 
  shorter period (\textless 500 d) and (\protect\subref{fig:fehpgt600}) longer period 
  (\textgreater  600 d) systems.  
  There is a paucity of iron-poor systems with high eccentricities.
  The short-period sample is dominantly supersolar, even more  
  so for systems having binary host stars.
Found using exoplanets.org. 
}
\label{fig:FehCorrEcc}  
\end{figure}

\begin{figure}[h] 
        \begin{subfigure}[ht]{0.46\textwidth}
                \centering
                \includegraphics[width=\textwidth]{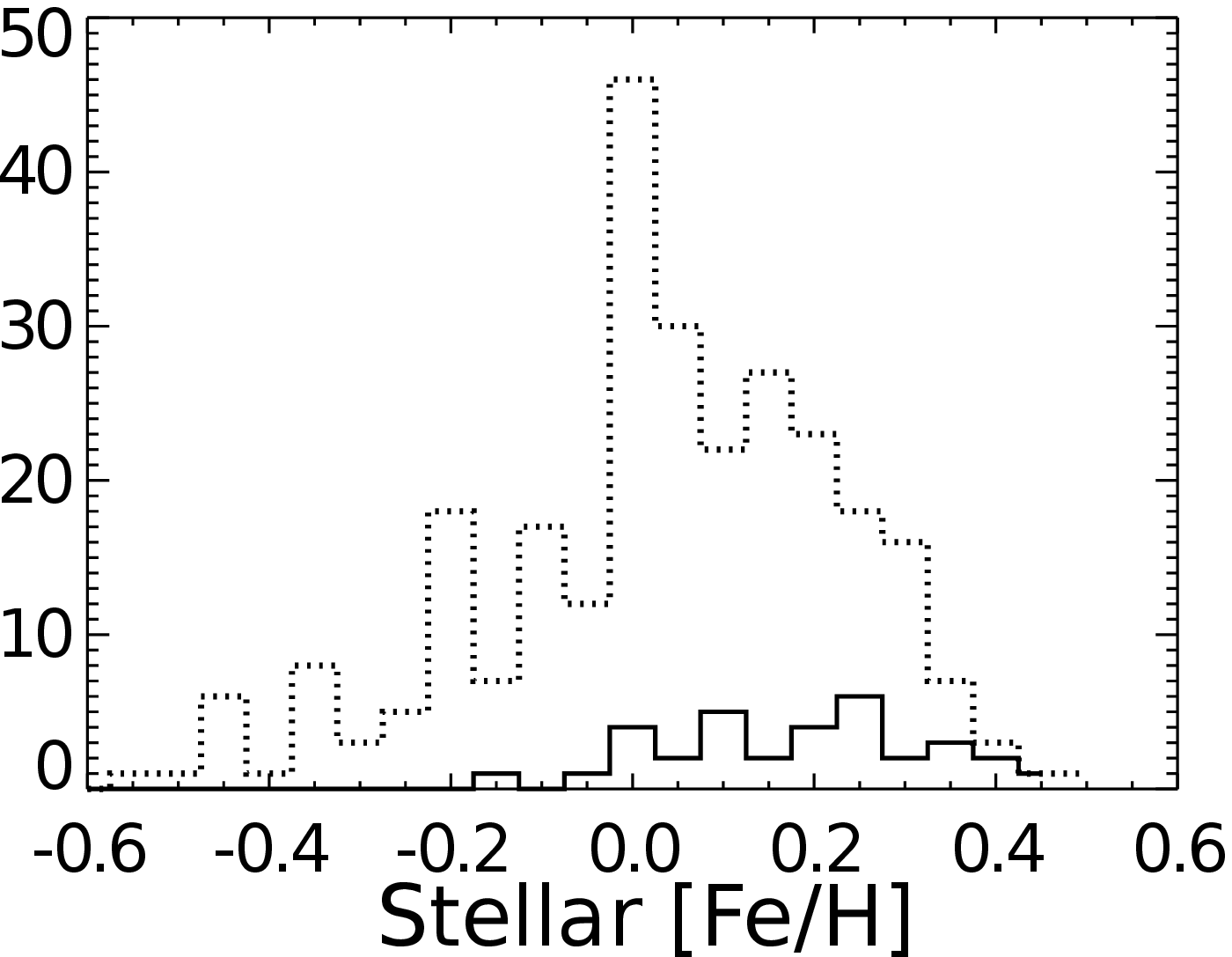}
                \caption{Counts of systems with periods \textless 500 days.}
                \label{fig:fehdivate35plt500}
        \end{subfigure}
        \begin{subfigure}[ht]{0.46\textwidth}
                \centering
                \includegraphics[width=\textwidth]{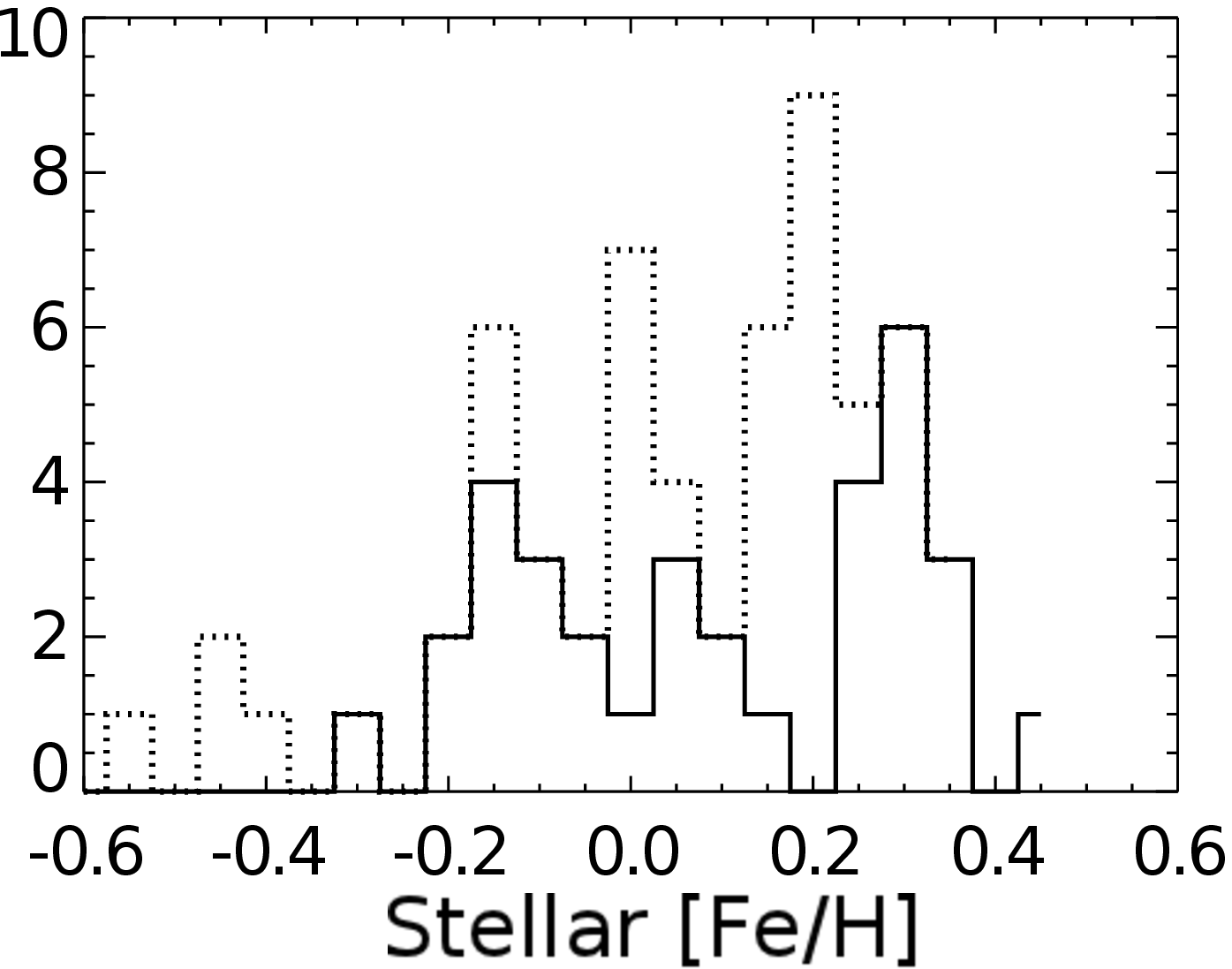}
                \caption{Counts of systems with periods \textgreater 600 days.}
                \label{fig:fehdivate35pgt600}
        \end{subfigure}
\caption{Histograms of [Fe/H]$_{\ast}$ for (\protect\subref{fig:fehdivate35plt500}) 
  shorter period (\textless 500 d) and (\protect\subref{fig:fehdivate35pgt600}) longer period 
  (\textgreater  600 d) systems, counting systems above 
  (dotted line) and below (solid line) 0.35 in each panel.
  The populations with higher eccentricity have clearly higher [Fe/H]$_{\ast}$ for systems with orbital periods less than 500 days, 
  but are still associated with moderately higher [Fe/H]$_{\ast}$ for systems with orbital periods greater than 600 days.
}
\label{fig:FehEccPersHist}  
\end{figure}

\section{Planets migrating towards and merging with star}    
\label{sec:migmerge}

One of the biggest surprises from finding enough planets to study them 
statistically has been how much planets continue to migrate after formation. 
Planets in eccentric orbits experience changing tides raised by the star's gravity
that cause the planet to migrate inwards (towards the star) and also circularizes 
the orbit. As  orbits migrate to shorter periods they become circular, almost always before the planet merges with 
star, leaving the short-period pileup of giant planets where the inward migration from tides on the planet typically ends. 
Planet formation leaves a short-period pileup of giant planets as well, 
so we do not know how much of the observed pileup is from formation or
is from more recent inward migration.

It was once thought that planets in the pileup could remain at the few-day 
periods of the pileup 
for the lifetime of the star, but work by \citet{jac09}, \citet{lev09} and others 
showed that most of the short period transiting planets found in 2009 are 
close enough to the star for the planet to raise tides on the star.
Most of these planets are destined to ``falling'', or inward migration 
ending in merger with the star.

The three-day pileup could either be entirely left over from formation
or it could be a product of ongoing migration, depending on whether 
planets migrate into and out of the pile up after formation. 
Only if the strength of stellar tides (raised by the planet) are not too
weak will planets move out of this pileup into the star.
The number of planets in the pileup affects the calculation of how many planets could be regularly 
migrating inwards on high eccentricity orbits that due to tides on the planet 
(raised by the star) bring planets inwards and reduce the eccentricity.
As the eccentricity is reduced, the inward migration stops, or only pauses 
if stellar tidal dissipation is strong enough to keep inward migration going. It is important to learn what the
contribution of inward migration to the pileup is.


We expect that planets migrate into the star through both slow and fast processes:
slow gradual migration going through a ``hot Jupiter'' phase, and fast high eccentricity
migration resulting from planet-planet scattering events.
Planets migration towards the star must result in many planets merging with the star, 
unless there exists impenetrable braking processes that stop all planets from
migrating into the star, resulting in the short period pileup.
We concern ourselves here with migration after dissipation of the disk, 
though note that disk migration can leave a pileup at short periods.


If these planets hosted by iron-poor stars represent planets which at other periods would have 
been associated with disruption of other planets into the star, but at these periods there were 
fewer inside planets to disrupt, then we might find a gap in the pileup of metal-rich stars. 
We find a pronounced gap in the number of iron rich stars, as shown in Figure~\ref{fig:PerHistFeHiVsLo}, 
in between two peaks of iron rich systems. 
This gap extends from less than 500 days to over 900 days, and is between peaks that 
go from 250 days to under 500 days, and from over 900 days going down gradually to over 2000 days.
In Figure~\ref{fig:PerHistFeHiVsLo-a-aswp} we show all systems, but in Figure~\ref{fig:PerHistFeHiVsLo-b-sswp}
we only show systems where no stellar companion has been found (SSWPs), because we show 
in Section~\ref{ssec:bswp} that systems for which the star has a stellar companion comprise  
a different population characterized by higher [Fe/H]$_{\ast}$ of the star. 
We made this cut because a high fraction of the systems in the gap were in BSWP systems,
and we see in Figure~\ref{fig:PerHistFeHiVsLo-b-sswp} that the gap is deeper though still in the same place. 

We find that the alternative hypothesis, that the iron-eccentricity correlation is due to higher 
iron abundance at formation, is only able to explain some but not all of these features.

Under the scenario that iron abundance leads to the scattering that leads to high eccentricity systems, 
the spike could be due to the snow-line pileup having such a strong tendency
to form crowded planets that leads to scattering even in iron-poor systems.
However, we find the gap between two peaks in the iron-rich distribution more difficult to explain.
It difficult to believe that higher initial crowding will lead to so much scattering that the initially
high population is depleted so thoroughly as to become this gap.

\section{Metallicity correlations with eccentricity and binarity}
\label{sec:correlation}  
Stellar [Fe/H]$_{\ast}$ and orbital eccentricity are very separate properties. 
FV05 cited the then apparent lack of correlation as evidence against the hypothesis
that the giant planet metallicity correlation was caused by pollution from the disk.
It is a surprise to find that [Fe/H]$_{\ast}$ and eccentricity are correlated, 
in separate but related patterns that is best explained by eccentric orbits coming
from events that raised the [Fe/H]$_{\ast}$.
The presence of a correlation of [Fe/H]$_{\ast}$ with eccentricity changes the debate
of whether the presence of planets is the result of primordial concentrations of iron,
or if the star has been polluted by factors related to the presence of planets.

The presence of a stellar companion might hinder the formation of planets, 
as well as increase perturbations leading to planet mergers with the star, 
so perhaps it is less surprising that we also find that the presence of a 
binary (or multiple) star companion to be correlated with higher [Fe/H]$_{\ast}$ in the planet host star.

More work will certainly have to be performed to determine how much of 
the [Fe/H]$_{\ast}$ is due to pollution and how much is from the initial [Fe/H].

%

\subsection{Stellar iron abundance correlations with eccentricity of orbit, with spike and dip} 
\label{ssec:FeEccCorr}
The eccentricities of planet systems with stellar  [Fe/H] ([Fe/H]$_{\ast}$) above and below solar 
represent two distinct populations. 
Figure~\ref{fig:EccVsPerSzFe} shows an obvious difference between all systems, 
Figure~\ref{fig:EccVsPerSzFe-All}, 
and iron-poor ([Fe/H]$_{\ast}$ $\le$ 0]) systems,  Figure~\ref{fig:EccVsPerSzFe-ltfe0}.
Iron-rich systems have a higher spread of eccentricity 
than  low [Fe/H]$_{\ast}$ systems at nearly all periods, 
except we find a narrow-period range ``spike'' of four subsolar [Fe/H]$_{\ast}$ systems with high eccentricity 
that is confined to a period range of 525 to 600 days.
The spike is also seen to fit into a semi-major axis range of 1.15 to 1.4 AU.
The highest eccentricities in both populations rises with period up to roughly the period range 
of the spike, but it appears that the maximum eccentricities may decline at longer periods.
The rise in eccentricity at short periods is expected due to tidal circularization, 
but the manner in which there is a different rise for the two populations was unexpected.
Though observational effects may produce the long period decline, 
observations should be blind to differences in iron abundance, 
so the appearance of different patterns that result from separating by [Fe/H]$_{\ast}$  suggests
that the patterns come from physical differences.

The summed occurrence distributions of iron-rich systems 
is also different from that of the iron-poor systems, Figure~\ref{fig:PerHistFeHiVsLo}.
At periods from under one year to a few years, iron-rich systems show a double peak separated by a 
prominent gap (\subref{fig:PerHistFeHiVsLo-a-aswp}) that goes from under 500 days to over 900 days. 
This gap becomes even more prominent when the 
systems with binary stars (BSWPs) are removed (\ref{fig:PerHistFeHiVsLo-b-sswp}).
In contrast, iron-poor systems have a single broad peak 
centered at roughly 1000 days.
The ``spike'' of iron-poor high-eccentricity planets thus corresponds to the inner part 
of the gap of iron-rich systems.





\begin{figure}[h] 
                \centering
                \includegraphics[width=0.9\textwidth]{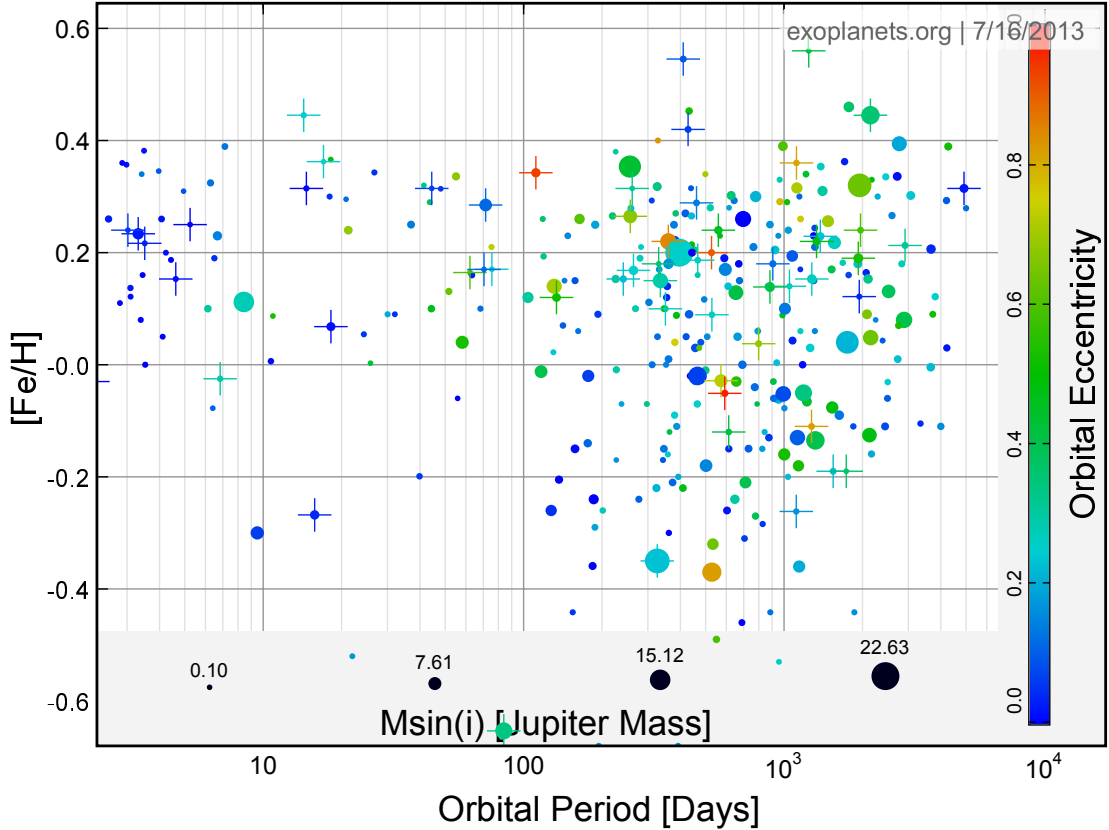}
\caption{Iron abundance versus period distribution, with crosses showing systems with binary companions,
  of systems with msini greater than 0.1 $M_{\rm J}$,  
  log g\textgreater 4.0, and 
  stellar effective temperature $T_{\rm eff}$ between 4500 and 6500 K.
  The marker size gives the planet mass, and the color gives the eccentricity. 
  The few highest [Fe/H]$_{\ast}$ 
  systems are in both of the double-peaks of 
  iron rich distribution. 
  }
\label{fig:FeVsP_showEcc}  
\end{figure}

The different nature of iron-rich versus iron-poor systems is further apparent when the systems are plotted by 
[Fe/H]$_{\ast}$ versus period as shown in Figure~\ref{fig:FeVsP_showEcc}, and 
[Fe/H]$_{\ast}$ versus eccentricity for periods shorter and longer than 
the spike, as in Figure~\ref{fig:FehCorrEcc}.
We see that the iron-poor population appears largely constrained to periods greater than
100 days. Most iron-poor systems with periods below 500 days have very low eccentricity. 
We note the presence of a few extreme high-iron systems, with [Fe/H]$_{\ast}$ \textgreater 0.38, 
that correspond to the both of the double peaks mentioned above. 
There is also a corresponding gap of extreme [Fe/H]$_{\ast}$ systems corresponding to the 
gap in occurrence distribution found above. Further work needs to be done to confirm 
if these features are simply due to more systems in the peak and fewer in the gap.
We doubt that the most iron-rich systems, especially those with  [Fe/H]$_{\ast}$ \textgreater 0.45, 
are high simply due to statistical scatter. We note that the two systems with  [Fe/H]$_{\ast}$ \textgreater 0.5
both consist of binary stellar systems. 
Of the five systems with  [Fe/H]$_{\ast}$ of 0.4 or higher with periods above 400 days, 
four have eccentricities above 0.35 and three are in binary systems --
none of the five is in a low eccentricity single star system.
Of the nine systems with 0.38 [Fe/H]$_{\ast}$ \textgreater 0.38, all but two have 
eccentricities above 0.35 -- one is a BSWP, and one has an eccentricity of 0.19,  
still moderately high.
Outside of this period range, there is only one system that has
[Fe/H]$_{\ast}$ \textgreater 0.40, HD 38529b which has a 14d period, and it is also a binary that has the unusually high 
eccentricity for its period of 0.24.
Clearly, high [Fe/H]$_{\ast}$ is associated with one or both of two parameters  
that could be associated with other planets being disrupted into the star.


\subsection{Evaluation of high and low [Fe/H]$_{\ast}$ systems being two populations} 
\label{ssec:EvalFeEccCorrv}
We show that star with planet (SWP) systems can be separated into iron-rich 
and iron-poor populations which are statistically different at short periods and at longer periods. 
We evaluate the significance of iron-rich and iron-poor planets being different populations
by first dividing star with planet (SWP) systems by period.               


A correlation was first found looking at  [Fe/H]$_{\ast}$ as a function of eccentricity, 
as shown in Figure~\ref{fig:FehCorrEcc}, by \citet{tay12b,tay13a} 
and \citet[hereafter DM13]{daw13}. DM13 presented that iron-rich and iron-poor systems 
are two populations, and showed that the three-day pileup of planets that had appeared to 
be largely missing in the Kepler data was in fact present in the iron-rich population.
Previously, only the short period correlation 

We show in Table~\ref{tab:CorrSummaryE35} 
that the level of eccentricity above which the stellar [Fe/H]$_{\ast}$ is higher 
is roughly 0.35 for planets with periods under 500 days, 
but increases to roughly 0.55 for planets with periods over 600 days.
The results of a T-means test for planets with periods less than 500 days
show that there are two separate populations above and below
eccentricity of 0.35, but that for planets with periods over 600
days, it can be seen with moderate confidence that there are
two populations when a division is taken at an eccentricity of 0.55.
For systems with periods under 500 days,
there may be a continuous rise in [Fe/H]$_{\ast}$ with
eccentricities above roughly 0.35, but it uncertain if this is significant.

The correlation of increasing [Fe/H]$_{\ast}$ with increasing eccentricity 
was found when looking for evidence for the infall of other planets
during the inward migration of hot Jupiters described in Section~\ref{sec:signatures}.
We were also seeking to explain the paucity of additional planets in hot Jupiter systems 
but not in hot Neptune systems \citep{fab12}, and the lack of an [Fe/H]$_{\ast}$ correlation
with smaller planets similar to the correlation that exists for giant planets \citep{buc12}
by hypothesizing that missing planets may have gone into the star
after their orbits were disrupted by the migration of the hot Jupiter.
We had hypothesized that planets with the highest eccentricity were the ones
to have most recently disrupted other planets, and might have higher
[Fe/H]$_{\ast}$ than stars with the shortest period planets because those stars
have had more time for the iron to dissipate down away from the surface.


\begin{deluxetable}{ cccccccccc } 
\tabletypesize{\footnotesize}  
\tablecolumns{10} 
\tablewidth{0pt}
 \tablecaption{ Correlations summary of regions for $P<500 d$ and $P> 600 d $. 
            \label{tab:CorrSummaryE35}                                                               }      
\tablehead{
  \colhead{Period}                                           & \colhead{Probability} 
  &\colhead{Number} &\colhead{$e<0.35$} &\colhead{$e<0.35$} &\colhead{} 
  &\colhead{Number} &\colhead{$e>0.35$} &\colhead{$e>0.35$} &\colhead{}      \\
  \colhead{(days)}                                           & \colhead{same} 
  &\colhead{SWP} &\colhead{Median} &\colhead{Mean} &\colhead{var-} 
  &\colhead{SWP} &\colhead{Median} &\colhead{Mean} &\colhead{var-}                                 \\
  \colhead{ region}                                         & \colhead{population} 
 &\colhead{$e<0.35$}  &\colhead{[Fe/H]$_{\ast}$} &\colhead{[Fe/H]$_{\ast}$} &\colhead{iance} 
 &\colhead{$e>0.35$} &\colhead{[Fe/H]$_{\ast}$} &\colhead{[Fe/H]$_{\ast}$} &\colhead{iance} 
}
\startdata    
\multicolumn{10}{c}{For SWP above and below an eccentricity of 0.35} \\
$< 500d $& $8 \times 10^{-3} $& 116 &0.010 & 0.071 & 0.05  & 29     & 0.21 & 0.19 &0.02  \\
$> 600d $& 0.40 & 60 & 0.12 & 0.08 & 0.05   & 33     & 0.13 & 0.12 & 0.043   \\
\hline \\
\multicolumn{10}{c}{For SWP above and below an eccentricity of 0.55} \\
$< 500d $& 0.014 &   135&0.11  & 0.088 & 0.05   & 10  & 0.21 & 0.20 &0.014  \\
$> 600d $& 0.035 &  84 & 0.090 & 0.080 & 0.05   & 9     & 0.29 & 0.24 & 0.03   \\
\enddata   
\end{deluxetable}

This difference between shorter and longer period planet systems 
favors an explaintion that an ongoing infall of planet material
results from the passage of a high eccentricity planet migrating through
the paths of other planets.
In addition to scattering other planets directly into the star, \citet{mat13} 
show that terrestrial planets may be caused to collide with other planets, 
and that these collisions may scatter material that could make its way to the stellar surface.
It more difficult to explain how 
higher formation rates in higher metallicity systems could cause this dependence.
The hypothesis that higher eccentricity planets are associated with more recent multiple planet migrations that could have polluted the star, likely acting as a form of positive feedback for systems that are more
likely to have planet scattering due to more crowded planet systems that are hypothesized to
be formed in systems with higher [Fe/H]$_{\ast}$ \citep{daw13}. 
\opt{optrplan}{I plan to see if convective mixing of pollution during the time an eccentric planet becomes a circularly orbiting ``hot Jupiter'' could account for the lack of correlations that have led others to reject the pollution hypotheses.}

Recently, \citet{daw13} independently found that those planets in higher
eccentricity orbits were correlated with the star having higher [Fe/H]$_{\ast}$, for which
they give the explanation that more giant planets are formed in more
metal rich systems, and that more crowded formation of giant planets leads
to more planet-planet scattering, resulting in more eccentric planets.
They also found that the pileup of short period giant planets (hot Jupiters)
is a feature of metal rich planets by recovering much of this pileup
in the Kepler data.

The best explanation may be that the two processes are complementary,
with both contributing to the stellar [Fe/H]$_{\ast}$-eccentric planet orbit correlation.
However, the occurrence of higher [Fe/H]$_{\ast}$ in BSWPs may be better
explained by increased accretion caused by the presence of a companion.
Star formation is not thought to be affected as much by high [Fe/H]$_{\ast}$
of the star-forming material, and stellar companions are more
disruptive to planet formation than helpful.

\citet{daw13} suggest that higher [Fe/H]$_{\ast}$ leads to more crowded giant planet systems,
leading to more planet-planet scattering that could create 
ongoing inward migration of giant planets.
They present the evidence of this flow of giant planets from Kepler planet candidates
that the so-called ``three-day'' pile up is a feature of metal rich but not of metal poor stars.
If metal-rich systems experience more planet scattering, then they would disrupt more
planets into the star. They note that even among iron rich systems that the pileup is still
smaller than in our solar neighborhood. Perhaps the hypothesis given by \citet{how12} 
that systems in the Kepler field tend to be older due to being further from the plane of the galaxy
explains the smaller pileup.

We find that stellar [Fe/H]$_{\ast}$ increases with eccentricity with a T-test confidence of 5\e{-5}
of being by chance for systems with
periods less than 500 days and $e>0.35$. 
For systems with periods greater than 600 days, there is a weaker correlation
stellar [Fe/H]$_{\ast}$ with eccentricity for systems with  $e>0.55$ with 
the probability of this being by chance given by a T-means test
as 8\%. There are only 11 systems in this range, so the less definite T-means test result can be expected.

For longer periods, it does appear that there is a weaker 
correlation, with the threshold moving from $e>0.35$ for period less than 200 day planets, 
to a threshold of $e>0.55$ for planets with period over 600 days. Whether planets in
the intermediate range of periods from 200 to 600 days makes a smooth transition is uncertain
due to the presence of a small number of systems with orbital periods between 500 and 600 days  
having planets with high eccentricity but stars with lower [Fe/H]$_{\ast}$.
It is difficult to explain why planets in the small period range of 500 to 600 days appear to 
be present at what appears to be a boundary between a higher and a lower level of
correlation of the stellar [Fe/H]$_{\ast}$ with eccentricity, but perhaps
when high eccentricity planets migrate through the ``planet desert'' 
(a region interior to the snow-line with fewer planets) there can be fewer planets to 
scatter into the star.

\subsection{Comparison of  [Fe/H]$_{\ast}$ of binary companions}
\label{ssec:binarycomparison}
\citet{lau97} suggest comparing the metallicity of binary companions as a test of pollution.
The binary system HD 80606/7 is remarkable for the high 
eccentricity of HD 80606b of $0.927 \pm 0.012$, 
and has high [Fe/H]$_{\ast}$ of $0.43 \pm 0.06$
while its companion, HD 80607 has high, but less high, [Fe/H]$_{\ast}$ of $0.38\pm 0.06$ \citep{nae01}.
If high [Fe/H]$_{\ast}$ leads to more scattering due to more crowded formation as
suggested by \citet{daw13}, 
but then the disruption of interior planets by the giant planet raised the value
further due to disruption into the star, then this could produce the result we see.
We urge more precise measurements of the metallicity difference between the two stars, 
and call for further measurements of other companions of planet hosting stars.
Comparisons of companion stars of similar masses has special value due to 
how the smaller companions that comprise the majority of binary companions
will have different convection layer thicknesses, making comparison of metallicity
requiring interpetation of the amount of mixing.     
We suggest that planet searches preferentially search on binary companions that
like HD 80606/7 have similar masses.
Comparing the metallicities of binaries will be made more difficult by how there is
often unexplained difference in binary star metallicities, 
though it may be common that planetary systems have had different
effects on each star. A search of the literature does not find a good discussion of
the spread of binary metallicity.

\subsection{Iron abundance correlated with presence of stellar binary companion}
\label{ssec:bswp}
We found that the mean [Fe/H]$_{\ast}$ for stars with binary companions that host planets (BSWPs)
is higher than the mean [Fe/H]$_{\ast}$ for SWPs when looking for further tests 
that companions to the star can increase accretion that raises the [Fe/H]$_{\ast}$
of a star's surface. 

\begin{figure}[h] 
        \begin{subfigure}[ht]{0.5\textwidth}   
                \centering
                \includegraphics[width=\textwidth]{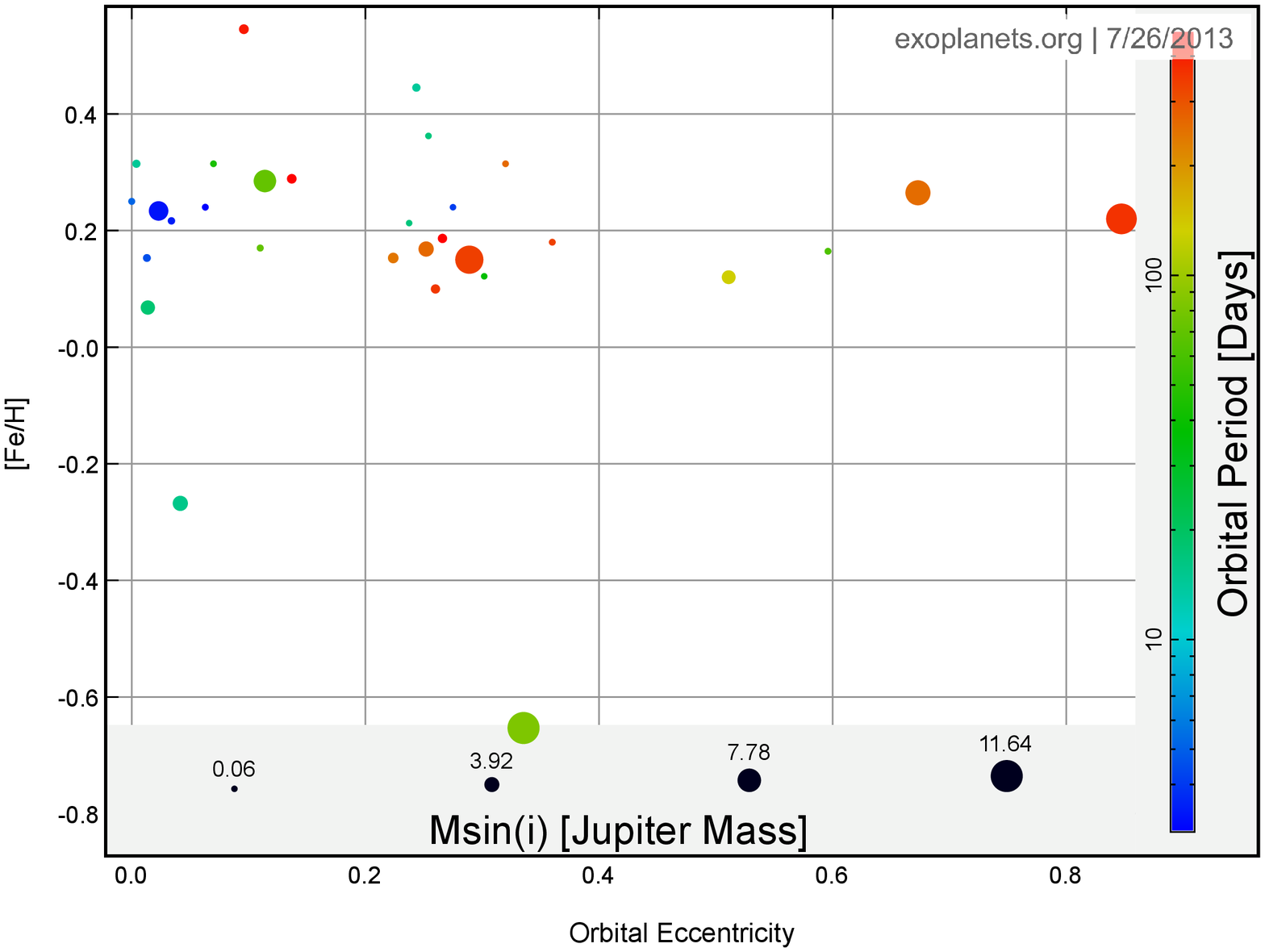}
                \caption{Periods less than 500 days.}
                \label{fig:fehbswpplt500}
        \end{subfigure}
        \begin{subfigure}[ht]{0.5\textwidth}   
                \centering
                \includegraphics[width=\textwidth]{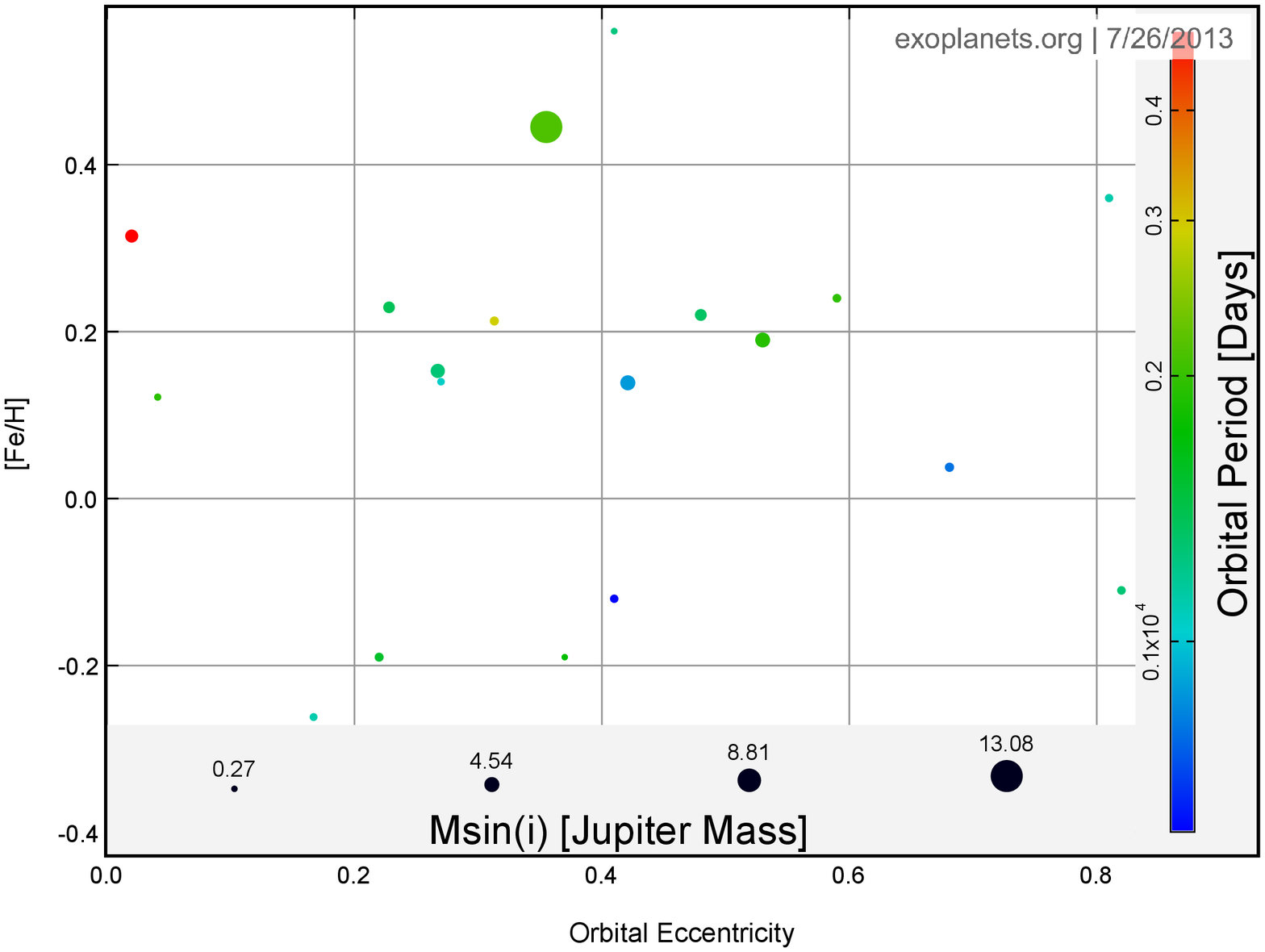}
                \caption{Periods more than 600 days.}
                \label{fig:fehbswppgt600}
        \end{subfigure}      
\caption{[Fe/H]$_{\ast}$ of BSWPs for periods below  (\protect\subref{fig:fehbswpplt500}) and       
  above  (\protect\subref{fig:fehbswppgt600}) the period range of the spike of high eccentricity iron-poor systems.
  The clustering of values closer to [Fe/H]$_{\ast}$ of 0.2 dex, compared to the wider and lower mean 
  distribution for all systems (Figure~\ref{fig:FehCorrEcc}), shows 
  that for stars having a binary companion that host planets, 
  [Fe/H]$_{\ast}$ is higher. 
  }
\label{fig:FehEccBSWP}  
\end{figure}

Stars with planets in less than a 500 day period that have a binary companion 
(BSWP) have higher [Fe/H]$_{\ast}$, 
with a 98\% probability of those planets having eccentricities below 0.35 
being a different population, according to a T-means test done as shown in 
Table~\ref{tab:bswpCorrSumm}, where we see that the mean [Fe/H]$_{\ast}$ is much higher.
 In Figure~\ref{fig:FehEccBSWP} we
compare the iron abundance of BSWPs and single SWPs (SSWPs). 
It appears that having a binary companion has a similar influence to the planet 
having higher eccentricity, as the difference between single stars
with planets (SSWP) and BSWP at high eccentricity is much smaller. 
We find that the presence of a binary companion significantly affects the 
correlation of the iron fraction with eccentricity, especially for 
systems with shorter orbital periods (under 500 days), 
such that stars with a binary companion hosting planets (BSWPs) rather generally 
have high [Fe/H]$_{\ast}$ from low to high eccentricities.
The statistics are poor, but BSWP with eccentricities below 0.02 may have lower [Fe/H]$_{\ast}$ than BSWPs with
higher eccentricity. 

We discovered this correlation when looking to see how many binaries would be
available to compare the [Fe/H]$_{\ast}$ of the SWP with the  [Fe/H]$_{\ast}$ of the companion without a planet.
We saw that for SWPs of periods below 500 days
 low eccentricities that BSWPs clearly had higher  [Fe/H]$_{\ast}$ than SSWPs, at a level
similar to the  [Fe/H]$_{\ast}$ of  SSWPs at high eccentricity. 
The  [Fe/H]$_{\ast}$ of BSWPs is much less dependent on eccentricity except [Fe/H]$_{\ast}$  may
be lower very close to zero but the statistics are too low to be certain.


We again find a small region in period space between 525 and 575 day periods
with a small number of systems 
having low stellar [Fe/H]$_{\ast}$ that may or may not be statistically significant,
but it looks apparent in the work of \citet{daw13} as well. 
We see only four or five SSWP and two BSWP systems in this region.
Perhaps there is less planet material inward of the snowline to scatter into the star.

Binary star formation is thought to occur through gravitational instability, 
so unless metallicity can play an unexpected role (perhaps by seeding), then 
this correlation represents a challenge to explaining higher iron abundance in
SWPs coming from a primordial higher iron concentration. 
It is also a challenge to explain how a binary star would increase accretion, 
but perhaps the additional orbital instability of the planets increases 
scattering. Another possibility is that a binary companion disrupts more 
of the original cloud into the star.

%
%


\begin{deluxetable}{ ccccccccccc }    
\tabletypesize{\footnotesize}  
\tablecolumns{11} 
\tablewidth{0pt}
 \tablecaption{ BSWP versus SSWP correlations summary of regions for $P<500 d$ and $P> 600 d $.
            \label{tab:bswpCorrSumm}                                                               }      
\tablehead{
  \colhead{Period} &\colhead{Eccentricity}  & \colhead{Probability} 
  &\colhead{Number} &\colhead{BSWP} &\colhead{BSWP} &\colhead{} 
&\colhead{Number}  &\colhead{SSWP} &\colhead{SSWP} &\colhead{}      \\
  \colhead{(days)} &\colhead{region}  & \colhead{same} 
  &\colhead{of} &\colhead{Median} &\colhead{Mean} &\colhead{var-} 
  &\colhead{of}  &\colhead{Median} &\colhead{Mean} &\colhead{var-}                                 \\
  \colhead{ region} &\colhead{} & \colhead{population} 
  &\colhead{BSWP} &\colhead{[Fe/H]$_{\ast}$} &\colhead{[Fe/H]$_{\ast}$} &\colhead{iance} 
  &\colhead{SSWP}&\colhead{[Fe/H]$_{\ast}$} &\colhead{[Fe/H]$_{\ast}$} &\colhead{iance} 
}
\startdata    
$< 500d $&$<0.35$ & 0.04& 25&0.22 & 0.18 & 0.05   & 116    & 0.100 & 0.071 &0.05 \\  
$< 500d $&$>0.35$ & 0.55 & 4 & 0.26 & 0.24 & 0.009 &29     & 0.21 & 0.19 & 0.02 \\
$> 600d $&$<0.35$ & 0.91& 8  & 0.15 & 0.090 & 0.042   & 60     & 0.12 & 0.081 & 0.047   \\
$> 600d $&$>0.35$ & 0.10 & 7 & 0.22 & 0.26 & 0.033   & 33       & 0.13 & 0.12 & 0.043   \\
\enddata   
\end{deluxetable}

\begin{figure}[h] 
                \centering
                \includegraphics[width=0.5\textwidth]{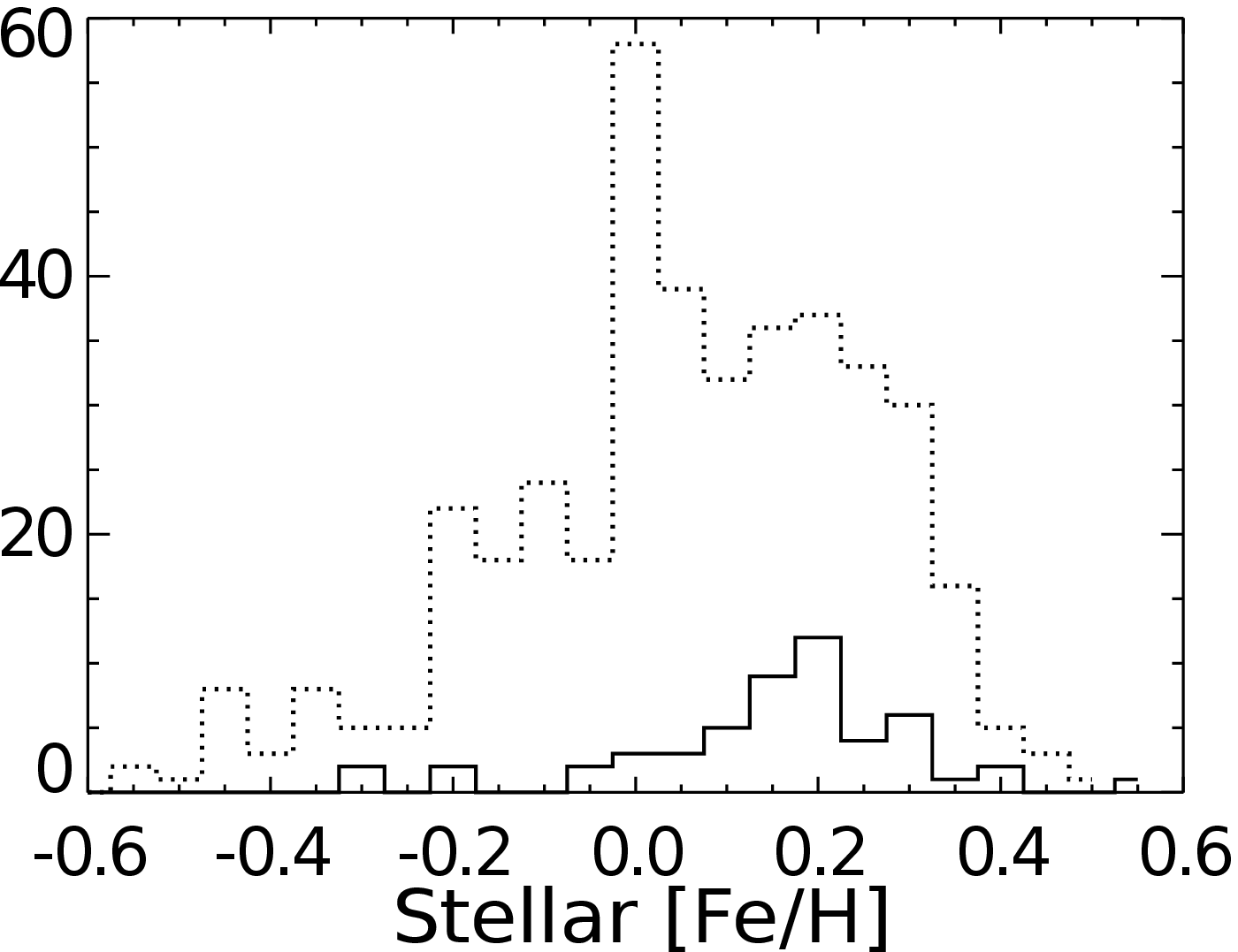}
\caption{Comparison of distribution of [Fe/H]$_{\ast}$ for SSWPs (dotted line) and BSWPs (solid line).
  This histogram shows that BSWPs are more iron-rich than SSWPs.
  The mean [Fe/H]$_{\ast}$ of BSWPs and SSWPs are 0.16 and 0.074, respectively.
  }
\label{fig:bivssinhefhist}  
\end{figure}

\subsection{Might raised [Fe/H]$_{\ast}$ be associated with planet disruption?}
\label{ssec:highestiron}
There is more structure in the  [Fe/H]$_{\ast}$ as a function of period, as shown in 
Figure~\ref{fig:FeVsP_showEcc}, than can be covered in this work.
There appear to be excesses of the highest and lowest [Fe/H]$_{\ast}$ systems in 
the regions of the two peaks and gaps, though it will take more work to fully
verify that these excesses are not due to higher numbers of high [Fe/H]$_{\ast}$
systems in the peaks, and the higher numbers of low [Fe/H]$_{\ast}$ systems in
the gap between the high [Fe/H]$_{\ast}$ peaks.
It is likely, however, that the presence of excess numbers of especially high or low 
[Fe/H]$_{\ast}$ systems will strengthen the result that the peaks and the gaps 
contain significant structure.

We specifically list these excesses: 
All systems with [Fe/H]$_{\ast}$ at or above 0.39 except one are in the same range as the 
two peaks in the occurrence numbers of high-iron systems shown in Figure~\ref{fig:PerHistFeHiVsLo}.
We count four high [Fe/H]$_{\ast}$  systems in the peak from 300 to under 500 days, 
and five high high [Fe/H]$_{\ast}$  systems if the longer period peak is considered to 
go from 900 to 3000 days, out of 248 objects.
The highest [Fe/H]$_{\ast}$ systems are overrepresent by BSWPs, with three of the nine
peaks systems belonging to systems with stellar companions, including the only 
two systems with [Fe/H]$_{\ast}$ greater than 0.5, even though there are 54 BSWPs
and 194 SSWPs among the 248 SWPs. The one system at shorter periods to have 
[Fe/H]$_{\ast}$ greater than 0.39 is a 10-day period BSWP.
In the gap between the two peaks, the highest [Fe/H]$_{\ast}$ is 0.315, yet has 
quite a few of the especially low [Fe/H]$_{\ast}$ systems.


\subsection{Lower lithium correlation}
\label{ssec:lithium}
Planet infall must be considered a possible factor that could cause the
lower lithium abundance in stars with planets \citep{isr09}.
Simulations show that when planets of several Jupiter masses ($M_J$) merge with solar mass stars, 
the planet's core survives a plunge past the convection zone (CZ) 
\citep{sha88,sie99b}, which would mix up the convection zone,
moving lithium rich surface layers down, and moving lithium poor 
lower layers upwards.
However, most systems will have only experienced the infall of a planet
less massive, and planets less than a few $M_J$ in mass will Roche-lobe overflow
above the star for what may be an extended period of time \citep{gu03}.
(\citet{spe11} report observing stars with ''false photospheres'' best explained by
the overflow of a planetary companion.)
This raises the question of how much of the surface lithium would be mixed deeper by  
the fall-in of Roche lobe overflow material into the star?

If migration of proto-hot Jupiters  scatters planets into the star more than the migration of smaller planets, this could explain why the Kepler mission has found that giant radii planets in multi-planet systems tend to be one of the further out planets, with the giant planets rarely the closest in planet \citep{fab12}. 


\section{Flow of planets evidenced by discrepancy in rates of inward planet migration} 
\label{sec:signatures} 

\begin{figure}
\centering
\includegraphics[width=10cm,clip]{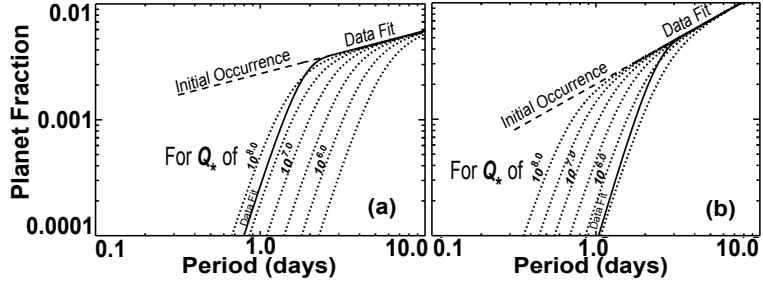}
\caption{Measured versus simulated occurrence distributions for short periods  
    for  (a) giant planets (summed for masses from $100$ to 2000 $M_\Earth$ 
    and radii from 8 to 32 $R_\Earth$) 
    and (b) medium planets (summed for masses from $10$ to 100 $M_\Earth$ and
    and radii from 4 to 8 $R_\Earth$).
    Fits to Kepler occurrence data from \citet{how12} are compared to calculated distributions for 
    a range of tidal dissipation strengths from weak to strong.  
    The left dotted-line curve is for a weak dissipation of $Q^{\prime}_{\ast}$ of $10^{8.5}$
    (remembering that $Q^{\prime}_{\ast}$ expresses the inverse of tidal dissipation, so lower values
    represent stronger dissipation), 
    with each next line for a factor of $10^{0.5}$ stronger dissipation, 
    up to a dissipation strength of $Q^{\prime}_{\ast}$ of $10^{6.5}$.
    The simulated distributions were calculated starting from an 
    initial occurrence distribution of no inner drop off, shown as a dashed line.
    Calculations were summed for a representative range of system ages. 
}
\label{fig:distLMo-c}       
\end{figure}

We find three signatures of inward planet migration that show that the shortest period planets are on their way to destruction by merger with the star: 1. The statistical distribution of giant and medium radii planets follows the power law expected for tidal migration into the star. 2. A higher amount of inward migration for giant planets than for medium and smaller radii planets could be indicated by the 
presence of more circular medium planet orbits compared to fewer uncircularized  
giant planet orbits
\citep{kan12} 
and, 3. The presence of a giant planet appears anti-correlated with other planets 
\citep{fab12}. We further explain the first reason in two different ways:
first looking at the measured distribution, and then by extrapolating to the future. 

The definitive determination for whether or not stellar dissipation is unexpectedly weak
will be the direct measurement of the whether planet periods are decreasing,
which should be  possible within a decade.  

\subsection{Power law shaped by inward migration}
\label{ssec:powerlawindex}
We use the fit to the occurrence distribution by 
\citet{how12}, who fit it to a ``Cutoff Power Law Model'' function, 
(``H12 fit'' hereafter).
 In this model, the function behaves like
a power law with exponent $\beta$ and
normalization $k_P$ for $P \gg P_0$, where $P_0$ is a characteristic period
below which the function has a dropoff towards a different power law:

\begin{align}\label{eq:howfiteq} 
\frac{\operatorname{d}\!{f(P)}}{\operatorname{d~}\!{ \log P}} =
  {k_P} {P^\beta}
\left[
  1-e^
    {-\left( 
      P/{P_0}
     \right)^\gamma }
\right] ,
\end{align} 

For the periods below the cutoff period $P_0$, the distribution $f(P)$
falls off more rapidly, becoming 
an exponential falloff with exponent $\beta + \gamma$ in the limit
$P \ll P_0$.


Table~\ref{tbl:howfitparsNoBeta} gives the best-fit parameters of the 
Power Cutoff Law model from H12, 
where, with large uncertainty, $\beta + \gamma$ is close to 13/3.

\begin{table}
\begin{center}
\caption{Best-fit parameters of ``Cutoff Power law'' Model, from H12.
  \label{tbl:howfitparsNoBeta}}
\begin{tabular}{crrrrr}
\tableline\tableline
$R_p (R_\Earth)$ & $k_p$ & $\beta + \gamma$ & $P_0$ (days) & $\gamma$ \\
\tableline
2-4 $R_\Earth$ &$0.064 \pm 0.040$ & $2.9 \pm 0.4$ &$7.0 \pm 1.9$  &$2.6 \pm 0.3$ \\
4-8 $R_\Earth$ &$0.0020 \pm 0.0012$& $4.8 \pm 1.3$ &$2.2 \pm 1.0$ &$4.0 \pm 1.2$ \\
8-32$R_\Earth$ &$0.0025 \pm 0.0015$ & $4.5 \pm 2.5$&$1.7 \pm 0.7$  &$4.1 \pm 2.5$ \\
2-32$R_\Earth$ &$0.035 \pm 0.023$ & $2.9 \pm 0.4$ &$4.8 \pm 1.6$  &$2.4 \pm 0.3$ \\
\tableline
\end{tabular}
\end{center}
\end{table}

\begin{figure}
\centering
\includegraphics[width=16.5cm,clip]{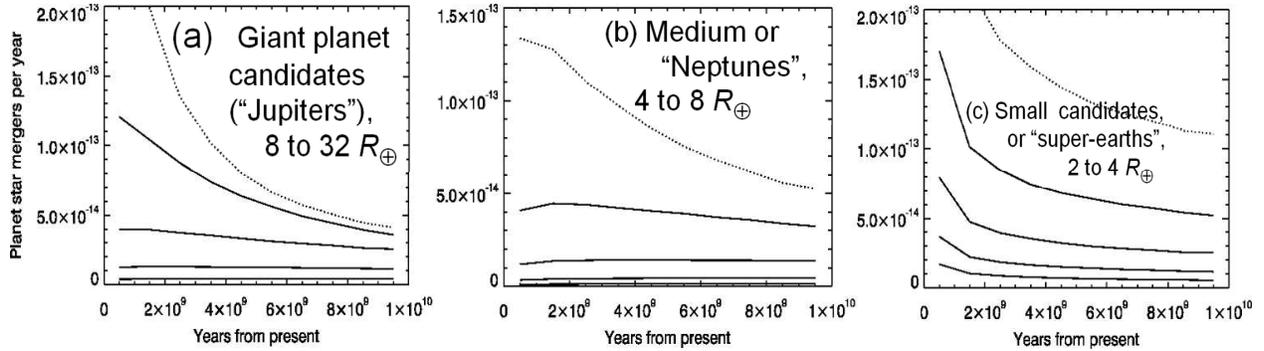}
\caption{Rates of calculated future planet mergers with host stars for three ranges of planet radii 
  based on the \citet[Howard et al. (2012); hereafter H12]{how12}   fit to the Kepler data, calculated for five values of 
  tidal dissipation strengths. Claims that tidal dissipation must be weak are based on
  arguing that the short period population cannot have a
  shortest-period distribution that will be depleted too quickly, 
  so that only curves representing a merger rate that are reasonably close to flat 
  represent the tidal dissipation strength of the stars hosting these planets. 
  (That is, for a tidal dissipation strength to be considered plausible, the 
  future inward migration rate based on that strength must only decrease 
  on the order of the ages of the stars.)
  This is only true if there are not a significant number of new planets 
  migrating in to replace the currently found shortest 
  period planets. We find that a substantial ongoing resupply of giant planets is not only reasonable,
  but better explains the discrepancy between the tidal dissipation strengths that this
  method gives for large and medium planets.
  We show curves for dissipation values of $Q^{\prime}_{\ast}$  
  values of $10^{6.5}$ (dotted, top), with each lower line representing a tidal dissipation 
  weaker by a factor of $10^{0.5}$, down to a strength of $Q^{\prime}_{\ast}$ of $10^{8.5}$. 
  }
\label{fig:rateforfit}
\end{figure}

We show what the occurrence distribution would be 
for a range of strengths of the tidal migration parameter $Q^{\prime}_{\ast}$, 
compared to occurrence distribution found by \citet[H12 hereafter]{how12} 
in  Figure~\ref{fig:distLMo-c}. 
The calculated distributions are found by using the 
tidal migration equations (TME) of \citet{jac09} 
operating on an initial condition of taking the long range part of the H12 fit
and making the simplification that there was no inner dropoff.
Running the TME for a distribution of system ages produces the dropoff
shown in  Figure~\ref{fig:distLMo-c}. We also performed the calculations
with an initial dropoff, and found that as long as the dropoff was not too close 
to the final dropoff, that the early infall was so fast that there was 
negligible difference with assuming no initial dropoff.

We first see that for giant and medium radii planets, that the 
hypothesis that the inner distribution is shaped by migration due to tides on the 
star matches the data well, by how close the power index 
of the dropoff in the H12 fit is 
to the value of 13/3 produced by the TME at zero 
eccentricity. The distribution of smaller planets (not shown) has a lower power index, 
indicating either a more primordial distribution or nonzero eccentricities. 
We next see that there is a discrepancy \citep{tay12a,tay12b} between the best fit of $Q^{\prime}_{\ast}$ 
to the data for giant and for medium planets: it appears that for giant planets
the turnoff is consistent with $Q^{\prime}_{\ast} {\gtrsim} 10^{7.5}$,
but for medium planets, a much lower value of $Q^{\prime}_{\ast} {\sim} 10^{6.0}$
is more reasonable. Given that binary star studies are more consistent with a lower 
value, we suggest that there could be a (greater) 
``Socrates flow'' \citep{soc12} of giant planets
that moves the measured occurrence distribution inward.

\subsection{Infall rate shows need for resupply of giant planets}
\label{ssec:infallresupply}
A way of evaluating the infall rate that does not rely on the ages of the system is to calculate 
the rate of future infall based on the current distribution, as shown in Figure~\ref{fig:rateforfit}.
We obtain this rate by summing the differential distribution of ``planet density'' 
that reaches the stellar radii as a function of time.
If there is no supply of new planets, the actual rate of infall should be decreasing only at the 
rate that the supply of planets decrease, and since no correlation of occurrence has yet been 
determined, the rate of infall should be constant or slowly decreasing. This view is equivalent 
to saying that the distribution outward of the drop-off should supply close to the same rate of 
infall as the shortest period distribution inwards of the cutoff. It can be seen in 
Figure~\ref{fig:rateforfit}a that for giant planets, tidal friction of 
$Q^{\prime}_{\ast} {\sim} 10^{7.0}$ 
would be reasonable for a slowly decreasing occurrence, and 
$Q^{\prime}_{\ast} {\sim} 10^{7.5}$ 
is consistent with occurrence remaining constant. But for medium planets, we see 
in Figure~\ref{fig:rateforfit}b that 
$Q^{\prime}_{\ast} {\sim} 10^{6.5}$ to $10^{7.0}$ 
is reasonable. This discrepancy can be explained by a continual resupply of giant planets, which could 	
keep the future rate of infall with $Q^{\prime}_{\ast} {\sim} 10^{6.5}$ 
to the either constant or slowly decreasing rate required.

\begin{figure}
\centering
        \begin{subfigure}[ht]{0.32\textwidth}
                \centering
                \includegraphics[width=\textwidth]{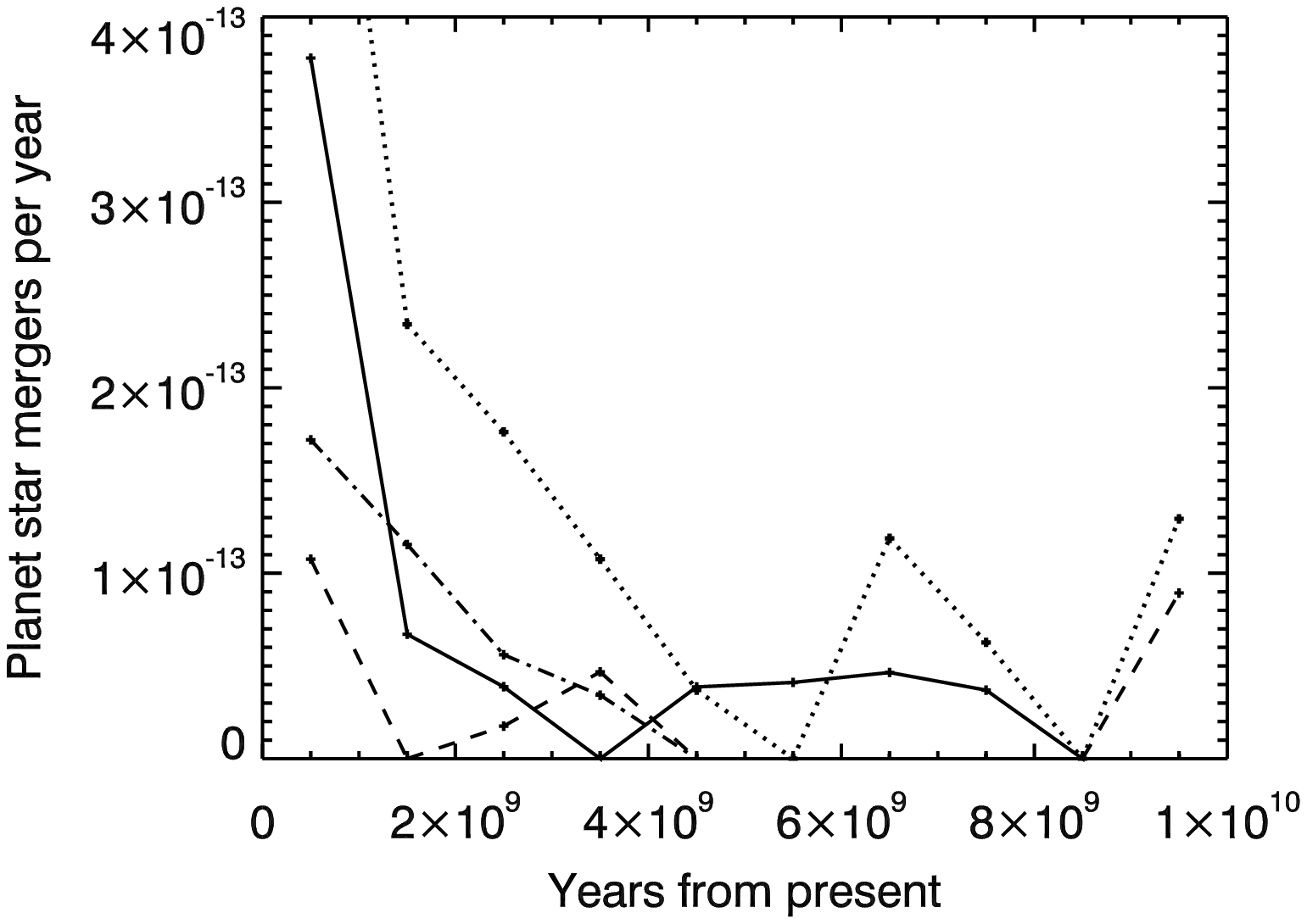}
                \caption{Infall rate of 8-16 $R_\Earth$ planets.}
                \label{fig:ratefordata_aJup}
        \end{subfigure}
        \begin{subfigure}[ht]{0.32\textwidth}
                \centering
                \includegraphics[width=\textwidth]{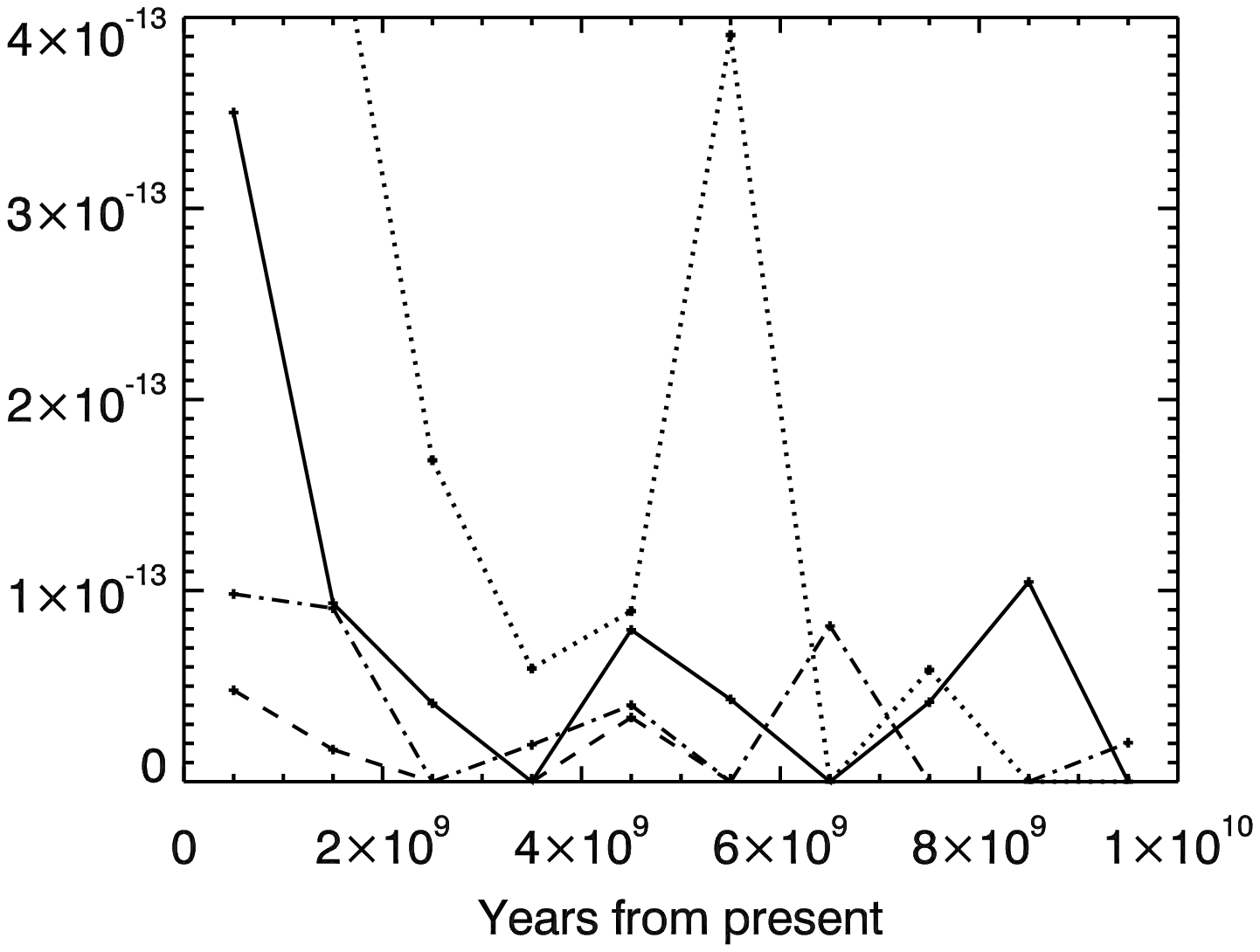}
                \caption{Infall rate of 4-8 $R_\Earth$ planets.}
                \label{fig:ratefordata_bNep}
        \end{subfigure}
        \begin{subfigure}[ht]{0.32\textwidth}
                \centering
                \includegraphics[width=\textwidth]{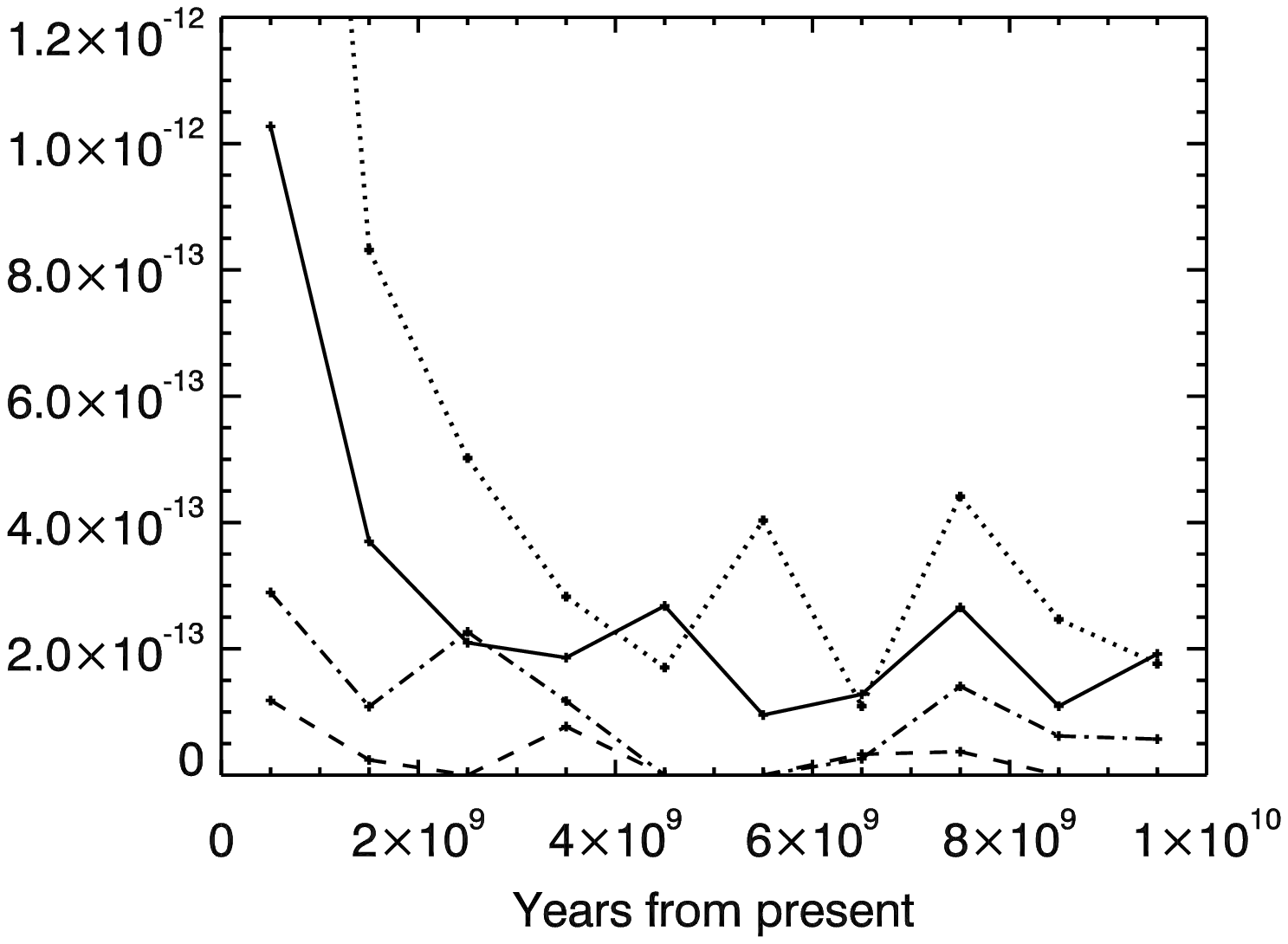}
                \caption{Infall rate of 2-4 $R_\Earth$ planets.}
                \label{fig:ratefordata_cSpe}
        \end{subfigure}
\caption{Future planet/star mergers calculated for actual Kepler candidates which are weighted 
  to give a rate that can be compared to the future merger rate presented in Figure~\ref{fig:rateforfit}.
  These are calculated for the same range of tidal dissipation strengths used Figure~\ref{fig:rateforfit}.
  Though the use of individual data points makes these results more noisy, the results are 
  consistent with those from using the fit. The appearance of peaks at  5 to 7 gigayears shows 
  that the tidal dissipation is unlikely to be as strong as $Q^{\prime}_{\ast}$ of $10^{6.5}$
  as this would result in an unphysical increase in merger rate when planets beyond the fall-off 
  migrate into the star.
  }
\label{fig:ratefordata} 
\end{figure}       

We show results from the same calculations for the future rates of infall based on 
the actual Kepler planet candidates rather than the fit in 
Figure~\ref{fig:ratefordata}. 
This data has been
 normalized to the 145,728 targets with Kepler magnitudes between 10.75 and 17.75
 \citep{bat12}, which, though different from the H12 normalization used in 
Figure~\ref{fig:rateforfit}, is reasonably close.
The noise in the actual data makes interpretation more difficult, 
but are consistent with the conclusion from modeling infall from the H12 fit 
that ongoing resupply of planets is the best 
explanation that support the rate of 
$Q^{\prime}_{\ast} {\sim} 10^{7.0}$ 
to explain the giant and medium planet distributions. A dissipation of 
$Q^{\prime}_{\ast} {\sim} 10^{6.5}$ 
 appears ruled out by how the larger number of planets beyond the drop-off it would arrive too quickly,
 producing an unphysical peak (at $\sim$5 Gyr).

If the infall of giant planets is due to a flow in addition to the rate calculated above, then
the required size of the flow can be considered to be the difference in the current rate of
infall and the rate at a far future time when the current shortest period planets have 
already fallen in. The current infall comes from the sharply declining part of the distribution
shown in Figure~\ref{fig:distLMo-c}, while the far future infall comes from the distribution
beyond the sharp infall. We can consider ``now" to be within the next 1 Gyr, and the
distant future when planets from the far distribution function have been falling in to
be 10 Gyr.

\begin{figure*}
\centering
\includegraphics[trim=0cm 0.2cm 0cm 10.6cm,width=12cm,clip=true]{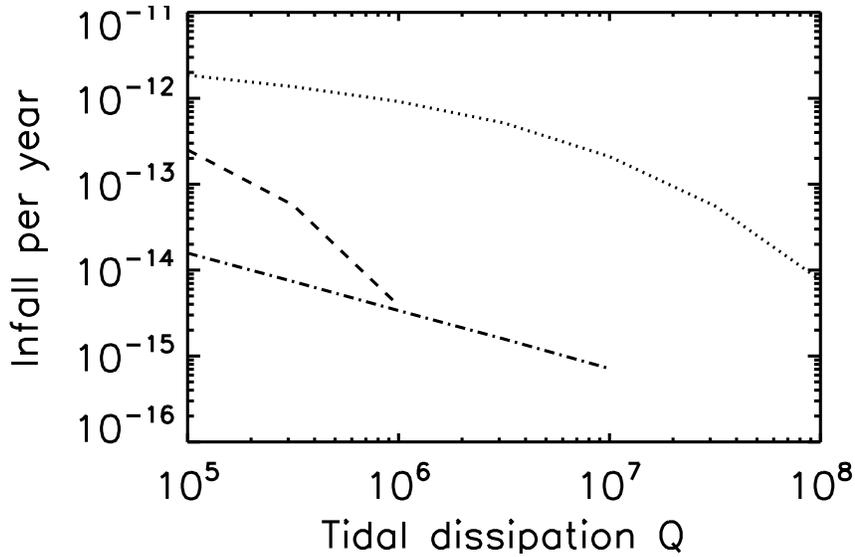} 
\caption{The rate that planets would need to migrate inwards to maintain the current 
population of shortest period planets, as a function of tidal dissipation strength $Q^{\prime}_{\ast}$. 
Rates are shown for the three radii ranges used in Figure~\ref{fig:rateforfit}, 
showing the required rates for giant planets (dotted) 
is higher than that for medium (dashed) and smaller (dot-dashed) planets.
These rates are obtained by taking the difference in planet infall
rate per year as a function of $Q^{\prime}_{\ast}$ 
between the flow at 1 and 10 Gyr, shown for three radii ranges. 
No line is shown for $Q^{\prime}_{\ast}$ values not needing a flow.
It takes only these rates of inward planet flow to maintain
a nearly constant future infall rate -- rates that we expect the planet
population should be able to supply.
}
\label{fig:rateVsQ}       
\end{figure*}

We calculate the rate of planet flow supplementary to 
tidal infall as a function of tidal dissipation strength required
as the difference in planet infall
rate per year as a function of $Q^{\prime}_{\ast}$ 
between the flow at 1 Gyr and 10 Gyr, which
we show  in Figure~\ref{fig:rateVsQ}.
It can be seen that the rate of planet infall would not unreasonably deplete the supply of planets
that would have to be scattered to provide this flow. Even for $Q^{\prime}_{\ast} {\sim} 10^{6.0}$,  
the flow would need to be less than only $10^{-12}$ planets per star per year.


\section{Discussion}
\label{sec:discussion}
If most eccentric planet migration is associated with disrupting other planets into the star
that raises [Fe/H]$_{\ast}$, then the region of the ``spike'' of low [Fe/H]$_{\ast}$ high eccentricity systems
could have arisen from these planets coming from where there were no other planets 
to be scattered into the star. It is provocative that these planets exist at a period range
likely interior to the snow line, where there may indeed be fewer planets to scatter inwards.
Might the lack of planets in the dip in the summed occurrence distribution also result from 
planets in this region not raising the star's [Fe/H] through causing planet infall?
Though some of these planets may have remained in the population of iron-poor planets,
we recognized that the dip represents too many ``missing'' planets for all of these 
planets to have been ``reassigned'' from the iron-rich population to the iron-poor population.


We were motivated to find the [Fe/H]$_{\ast}$ correlation with eccentricity by the idea that the
iron might mix into the star during the time that giant planets migrate from high eccentricity
to circular.
This could also explain that high  [Fe/H]$_{\ast}$ is not correlated with spin orbit misalignment. 
This would not explain that BSWPs have high  [Fe/H]$_{\ast}$ at most eccentricities.
Unless binary star formation is also increased by higher iron concentrations, 
the formation scenario also has trouble explaining this, because
the gravitational instability model of star formation is not thought to be influenced by
the metallicity.

We advocate including binary stars in planet searches both to get better statistics of
the metallicity of BSWPs, and also to better compare whether the companions of 
SWGPs (stars with giant plants) have different
metallicities from binaries with no giant planets. 
We also call for more theoretical work
seeking to separate the effects of having a stellar companion.


A list of the evidence for hypothesis that higher [Fe/H]$_{\ast}$ is at least 
partially from disrupted planets, due to an ongoing flow, and 
supposing tidal dissipation to not be weak, includes:
\begin{itemize}
  \item Overabundance of ``singleton'' hot Jupiters, e.g. systems with only one planet, 
  compared to predictions from planet formation models \citep{han13}. Since these
  planets are expected to form beyond the snow line, yet planets within the snowline are common,
  this suggests that there are missing planets, many of which must have ended up infalling into the star.
  \item Spin-orbit misalignment also gives evidence that planet migration initiated by scattering (DM13)
  must have occurred. The lack of correlation of between
  spin-orbit misalignment and metallicity was one factor that led us to consider that the measured metallicity
  may have been reduced during migration from a high eccentricity orbit to circularized short period orbits.
  \item Excess of giant planets compared to medium planets at shortest period.  
  \item Correlations of [Fe/H]$_{\ast}$ with eccentricity.
  \item The smaller pileup in the Kepler field than in the solar neighborhood \citep{how12}.
  may have been due to the pileup eroding with age as suggested by \citet{how12}. 
  If the association of the pileup found by \citet{daw13} is, as they suggest, associated with greater
  scattering, then such giant planet migration started by scattering would likely cause planet infall.
\end{itemize}

\section{Further work}
\label{sec:FurtherWork}
The next work will be to investigate what quantities of movement of  
planets and planetary material as a function of time could reasonably account for
these observations. We say we need a "by the numbers" study of how many 
planets are scattered and migrating at high eccentricity, followed by how many 
planets are in the pileup and then undergo short period migration into the star,
including dividing up how many planets collide versus undergo longer-term
Roche lobe overflow (RLO) \citep{tay10}, and how much material gets into stars how fast.
Finally, we will need to study the star, to see how much planet material is deposited how fast
and how long will this material last in the CZ before either convection or movement
through ``magnetic fingers'' \citep{vau04} takes the material below the CZ where the pollution
is no longer observed.

Resolving how much of these higher levels of metallicity in stars is from planet infall
versus being a signature of higher initial metallacity could be addressed by several studies:

\begin{itemize}
  \item 
Compare the time scales of mixing of pollution in the CZ with how long the times of migration from
eccentric planets to close circular planets. Damping of eccentricity is more dependent on the 
tidal dissipation  $Q^{\prime}_{p}$ value of the planet than  $Q^{\prime}_{\ast}$ of the star.
It is expected that $Q^{\prime}_{p}$ is not as weak as for the star.
  \item   
Compare the numbers of planets in pileup to what population necessary to supply an infalling flow
for a long enough time to explain the observed distribution.
See if this explains the remaining reduction in size between the pileup found 
for metal-rich planets by \citet{daw13} in the Kepler field and the pileup in the solar neighborhood.
How much ongoing planet-planet scattering will be required in the populations beyond the snowline?
How observable will a decline in planet populations be?
  \item
What is the rate of planet merger with the star as a function of time that is the best fit to the 
observations? Does this rate decay exponentially, and with what power index?
Is it possible to observe a decrease in the occurrence distribution with the ages of stars?
  \item 
Compare the [Fe/H]$_{\ast}$ of stars with with multiplicity of planet systems. 
See if stars with multiple planets might have lower [Fe/H]$_{\ast}$ because their
planets have not been scattered into the star. We make a tentative prediction that 
less disrupted planet systems might show lower [Fe/H]$_{\ast}$ --
tentative because more planets do not necessarily preclude planet scattering.
  \item
Astereoseismological studies to determine whether higher levels of metallicity are present 
throughout the star.
  \item 
We predict stronger tidal dissipation in the star, such that it should not be too many years
before a decrease in the periods of some planets should become measurable. 
\end{itemize}

\section{Conclusions}
\label{sec:conclusions}

We present evidence for continuing planet migration into the star, 
and a correlation of stellar iron abundance with three different populations
of parameters of planet orbits.

Beyond three day periods, there are more medium radii planets than giant radii planets
in the distribution of Kepler planet candidates, but below periods of two or three
days, there are more giant planets. This corresponds to the unexpected
presence of planets of a Jupiter mass or more at the shortest periods being
found by ground-based surveys. Since more massive planets migrate faster into
the star, we show how the shape of the occurrence distributions is better
explained by ongoing migration of giant planets into the star, rather
than saying that finding too many planets in the last short period before they
disappear into the star is evidence of weaker tidal migration in stars
for giant planet companions than for stellar companions.

This lead us to find new correlations between planetary systems that might
have had migration with stellar iron abundance. 
We present new patterns in planet parameters as a function of [Fe/H]$_{\ast}$ 
that demonstrate that planets of iron-rich and iron-poor systems 
constitute two different populations, with planets of multiple star systems
constituting a third population. 
We were investigating the hypothesis that an eccentric giant planet would be associated with
the most recent disruptions of an originally multi-planetary system 
that likely would have perturbed other planets into the star.

Different patterns can support either that higher iron abundance increases formation,
or that higher iron abundance is associated with planet pollution.
The shortest period distribution of giant planets shows that
unless there is a discrepancy in the stellar tidal migration strength as a function of the mass in the 
giant to medium planet mass range, the 
infall of giant planets remains an ongoing process. 
However, it is natural that the presence of a stellar companion would raise the 
metallicity threshold of planet production, so it is likely that both initial
iron abundance and planet pollution are factors in causing the giant-planet/metallicity
correlation.

Whether the higher iron in stars with stellar companions is less easy to explain in this model.

Having found evidence that the shortest period giant planets are in fact falling into the star,
we had considered whether these planets might have caused other planets to infall during their migration,
but we found that the  stars hosting the shortest period planets do not have higher [Fe/H]$_{\ast}$, so we 
considered whether the inward of these planets had been going on for enough time for pollution
to have been mixed away from the stellar surface. 
We considered whether the highest eccentricity planets might be the planets associated with
the most recent and the largest amount of planet scattering 
resulting in planet material being scattered to the stellar surface, and we considered that 
perhaps this trend would be strongest for shorter period planets as
longer period planets might have much slower migration and also might be less likely to 
scatter a planet or planet material into the star.

We find that the correlation of [Fe/H]$_{\ast}$ with eccentricity that decreases with longer period to fit this picture well.
Though we emphasize that giant planet migration may be ongoing, we see a 
a narrow period region that might correspond to planets migrating through a 
``desert'' region with fewer planets to scatter.

The presence of higher [Fe/H]$_{\ast}$ in stars with planets that also have a binary companion, with less 
correlation with eccentricity except at the lowest eccentricities, is not so easily explained by our model
nor is it easy to explain for the model of higher primordial abundance leading to higher rates of
giant planet formation. It prompts us to ask if we should consider whether 
high metallicity increases the rate of formation
of stellar companions? It is not implausible that the formation of binary stars is seeded by core accretion,
but this is not part of the standard gravitational collapse model of star formation.
It is essential further work to study the metallicities of multiple stars to see
how the presence of stellar companions affects metallicity.


New techniques have enabled the ability to study planets in their final epoch as they
migrate into and merge with their host stars. The destruction of planets into the star is certain to be 
one of the best opportunities to study planets, as they 
are taken apart by the gravitational forces of or suddenly destroyed by the outright collisions with their stars.
The merger of planets with stars will surely be among the most fascinating solar system events. 


The interplay  between the relationship of metallicity with formation and pollution
is likely to be complex, and require a better understanding of the evolution of
both stars and planets to fully explain.
We are lucky that there are many observables that can result from 
formation and planet migration, such that we can hope to disentangle which
processes produces which effects.

\section{Acknowledgments}
\label{sec:acknowledgments}
We gratefully acknowledge readings of the manuscript by 
Rosemary Mardling, Wen-Ping Chen,
Benjamin Bromley, Edward Guinan, Peter McCullough, and Daniel Fabrycky.





\end{document}